%% file: iclr2026_conference.tex
\documentclass{article} % For LaTeX2e
\usepackage{iclr2026_conference,times}

\input{math_commands.tex}

\usepackage{hyperref}
\usepackage{url}

\usepackage[utf8]{inputenc} % allow utf-8 input
\usepackage[T1]{fontenc}    % use 8-bit T1 fonts
\usepackage{hyperref}       % hyperlinks
\usepackage{enumitem}
\usepackage{url}            % simple URL typesetting
\usepackage{booktabs}       % professional-quality tables
\usepackage{amsmath}
\usepackage{amsfonts}       % blackboard math symbols
\usepackage{nicefrac}       % compact symbols for 1/2, etc.
\usepackage{microtype}      % microtypography
\usepackage{xcolor}         % colors
\usepackage{amssymb}
\usepackage{multirow}
\usepackage{adjustbox}
\usepackage{listings}
\usepackage{wrapfig}
\usepackage{graphicx}
\usepackage{subcaption}

\usepackage[most]{tcolorbox}   % in your preamble
\usepackage{tcolorbox}
\usepackage{float}
\newtcolorbox{gptprompt}{
  colback=gray!5,
  colframe=gray!50,
  boxrule=0.4pt,
  arc=2pt,
  left=4pt,right=4pt,top=4pt,bottom=4pt,
  fontupper=\footnotesize,
    before upper={%
    \setlength{\parindent}{0pt}%
    \setlength{\parskip}{4pt}%             % ← paragraph spacing
    \raggedright
  },
  breakable,                  % allow page breaks
  title=Prompt,                % optional title
}

\newtcolorbox{example}[2][]{%  #1 = extra keys (optional), #2 = title (mandatory)
  colback=blue!5,
  colframe=blue!50,
  boxrule=0.4pt,
  arc=2pt,
  left=3pt,right=3pt,top=3pt,bottom=3pt,
  fontupper=\footnotesize,
  before upper={%
    \setlength{\parindent}{0pt}%
    \setlength{\parskip}{2pt}%
    \raggedright
  },
  breakable,
  title={#2},   % use the second argument as box title
  #1            % allow further overrides
}

\newtcolorbox{question}[2][]{%  #1 = extra keys (optional), #2 = title (mandatory)
  colback=gray!5,
  colframe=gray!50,
  boxrule=0.4pt,
  arc=2pt,
  left=3pt,right=3pt,top=3pt,bottom=3pt,
  fontupper=\footnotesize,
  before upper={%
    \setlength{\parindent}{0pt}%
    \setlength{\parskip}{2pt}%
    \raggedright
  },
  breakable,
  title={#2},   % use the second argument as box title
  #1            % allow further overrides
}

\usepackage{pifont}
\newcommand{\cmark}{\ding{51}} % Check mark
\newcommand{\xmark}{\ding{55}} % Cross mark

\newcommand{\header}[1]{\noindent{\bf #1}}

\newcommand{\name}{{SmartDJ}}

\usepackage[textsize=tiny,textwidth=3.3cm]{todonotes}

\title{Guiding Audio Editing with Audio\\ Language Model}

\author{
    Zitong Lan\hspace{15pt} Yiduo Hao\hspace{15pt} Mingmin Zhao \\
    \textnormal{University of Pennsylvania} \\
}

\iclrfinalcopy % Uncomment for camera-ready version, but NOT for submission.
\begin{document}

\maketitle

\input{sections/00_abstract}
\input{sections/01_intro}
\input{sections/02_related}
\input{sections/03_method}
\input{sections/04_experiments}

\input{sections/05_discussion}

\bibliographystyle{iclr2026_conference}
\bibliography{iclr2026_conference}

\newpage
\input{sections/06_appendix}

\end{document}

%% file: math_commands.tex
\usepackage{amsmath,amsfonts,bm}

\def\eqref#1{equation~\ref{#1}}
\def\1{\bm{1}}

\DeclareMathAlphabet{\mathsfit}{\encodingdefault}{\sfdefault}{m}{sl}
\SetMathAlphabet{\mathsfit}{bold}{\encodingdefault}{\sfdefault}{bx}{n}

%% file: sections/00_abstract.tex
\vspace{-15pt}
\begin{abstract}
\vspace{-5pt}

Audio editing plays a central role in VR/AR immersion, virtual conferencing, sound design, and other interactive media. 
However, recent generative audio editing models depend on template-like instruction formats and are restricted to mono-channel audio.
These models fail to deal with declarative audio editing, where the user declares what the desired outcome should be, while leaving the details of editing operations to the system.
We introduce \name{}, a novel framework for stereo audio editing that combines the reasoning capability of audio language models with the generative power of latent diffusion. 
Given a high-level instruction, \name{} decomposes it into a sequence of atomic edit operations, such as adding, removing, or spatially relocating events.
These operations are then executed by a diffusion model trained to manipulate stereo audio. 
To support this, we design a data synthesis pipeline that produces paired examples of high-level instructions, atomic edit operations, and audios before and after each edit operation. 
Experiments demonstrate that \name{} achieves superior perceptual quality, spatial realism, and semantic alignment compared to prior audio editing methods. 
Demos are available at \href{https://zitonglan.github.io/project/smartdj/smartdj.html}{project page}.

\end{abstract}

%% file: sections/01_intro.tex
\vspace{-5pt}
\section{Introduction}
\vspace{-5pt}

Imagine you recorded a forest on a rainy day and wanted to transform it into the soundscape of a sunny forest. Achieving this transformation requires many edits: removing mismatched elements like rainfall and adding new effects such as bright leaf rustling. Traditionally, audio editing is a {\it procedural process}: the user specifies how to achieve the goal, step by step, by removing rain, layering new samples, adjusting gain, and so on.
Yet in practice, users would prefer just to issue a single, high-level instruction, e.g., {\it ``make this sound like a sunny day''}, and rely on an intelligent audio editor to decide what edits are needed. We call this {\it declarative editing}: the user declares {\it what} the desired outcome should be, while leaving the {\it how}, i.e., the sequence of operations, to the system.
Such a declarative paradigm would unlock a wide range of applications in VR/AR, gaming, cinematic post-production, and beyond, where designing and modifying audio scenes is central to immersive experiences.

Recent advances in deep generative models have opened exciting possibilities for text-to-audio generation and language-driven editing~\cite{evans2024fast, evans2024stable, hai2024ezaudio, heydari2024immersediffusion, huang2023make, liu2023audioldm, liu2024audioldm, jia2024audioeditor, liang2024wavcraft, wang2023audit, xu2024prompt}. 
However, existing audio editors remain limited in two key aspects. 
First, they rely on templated instructions such as “add the sound of birds” or “remove the sound of rain”, which restricts their ability to handle high-level or abstract user instructions. 
Second, they operate only on monaural (single-channel) audio, discarding spatial features such as interaural time and level differences, which are primary cues for spatial hearing. 
Without these cues, even semantically correct edits sound flat and fail to deliver immersive experiences.

In contrast, declarative editing requires an editor to bridge from high-level goals to detailed operations automatically.
When a user issues instructions such as {\it “place me in a concert hall”} or {\it “have this audio in a library”}, the system must reason about which sounds to remove, which to preserve, how to adjust loudness, when to introduce new events, and how to shift spatial position.
Current audio editors based on diffusion models cannot achieve this, as they lack the reasoning capacity to interpret these high-level instructions.
On the other hand, language models alone are also insufficient: while they can parse the user’s request, they lack grounding in the audio itself, making it impossible to decide which sound elements should be suppressed, emphasized, or retained.
This gap highlights the need for a framework that jointly reasons over language and audio, combining natural-language instruction understanding with audio-aware analysis and editing.

\begin{figure}[t]
    \centering
    \vspace{-8pt}
    \includegraphics[width=0.7\linewidth]{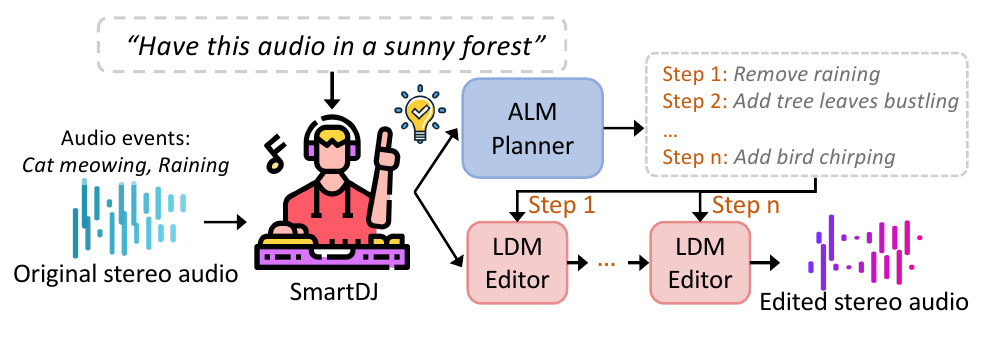}
    \vspace{-8pt}
    \caption{
    \name{} has an Audio Language Model (ALM) that acts as an edit planner to decompose the instructions into atomic steps, guiding the Latent Diffusion Model (LDM) editor to produce high-quality edited audio.
    }
    \vspace{-5pt}
    \label{fig:teaser}
    \vspace{-8pt}
\end{figure}

In this work, we present \name{}, the first framework for declarative audio editing by introducing Audio Language Models (ALM) into the audio editing loop.
This approach is motivated by the recent success of MLLMs in multimodal grounding and reasoning~\cite{chu2023qwen, ghosh2025audio, kong2024audio, liu2023llava, li2023multimodalfoundationmodels, zheng2025scalable}.
As illustrate in Fig.~\ref{fig:teaser}, our key idea is to let the ALM act as a {\em planner}: it perceives the original audio while interpreting the user’s high-level instruction and decomposes it into {\it a sequence of atomic edit operations}, such as adding or removing a sound, shifting direction, or adjusting volume, etc.
These atomic edit operations are then executed sequentially by a conditional Latent Diffusion Model (LDM) as an audio {\em editor}, realizing the user’s high-level goal.
This separation of planning and editing transforms audio editing from a procedural task into a declarative one.
Moreover, because the intermediate representation is expressed in natural language, users can easily inspect, refine, or override the planned edits interactively.

Training a system to perform declarative editing introduces a fundamental challenge: it requires paired examples of {\it high-level instructions}, their corresponding {\it atomic edit sequences}, and the {\it before-and-after edited audio}. 
Such data is rarely available, as natural soundscapes are complex and difficult to edit at scale.
To address this, we construct a scalable data generation pipeline that provides {\it controllable audio scenes}. Each scene is assembled from independently parameterized sound events with attributes such as direction and loudness.
Within this pipeline, an off-the-shelf LLM acts as the {\em designer}, producing diverse high-level editing instructions and the corresponding atomic operations. Audio signal processing then serves as the {\em composer}, rendering each operation by adjusting or mixing sound events to produce the corresponding audio after every edit.
Together, this designer–composer pipeline mirrors the planner–editor structure of our model. It produces a large-scale corpus of realistic editing operations, providing the supervision needed to train and evaluate our model.

Experimental results show that \name{} delivers superior editing quality and better alignment with high-level user instructions from both subjective metrics and human evaluations.
Our LDM also outperforms existing baselines for audio editing.
Ablation studies show that our ALM can effectively reason about and decompose high-level user instructions into a sequence of editing actions.

In summary, our main contributions are as follows:
\begin{itemize}[leftmargin=*,itemsep=0pt,topsep=-4pt]
\item We introduce \name{}, the first stereo audio editor capable of interpreting high-level user instructions with an ALM and executing them as precise atomic edit operations through an LDM.
\item We introduce the first scalable pipeline for generating editable stereo audio scenes, combining high-level instructions with controllable events to enable reasoning-based audio editors.
\item  We conduct extensive experiments and user studies with different baseline methods and demonstrate that \name{} achieves the highest editing quality for both objective and subjective metrics.
\end{itemize}

%% file: sections/02_related.tex
\vspace{-2pt}
\section{Related work}
\vspace{-5pt}
\header{Audio generation and editing.}
With the current advances in deep generative models, lots of methods have achieved high-quality audio generation from text and multi-modal conditions~\cite{chen2024video, evans2024fast, hai2024ezaudio, huang2023make,liu2023audioldm,liu2024audioldm, wang2023audit}.
Recently, spatial audio generation has attracted more attention~\cite{evans2024fast, heydari2024immersediffusion,liu2025omniaudio, sun2024both}, providing more realistic and immersive listening experiences.
Parallel to these generation efforts, text-guided audio editing also emerged as a powerful tool for modifying existing audio recordings.
Audit~\cite{wang2023audit} introduced an end-to-end diffusion model conditioned on both the input audio and simple, structured text commands, but its reliance on fixed editing templates limits flexibility to interpret high-level user prompts.
WavCraft ~\cite{liang2024wavcraft} leverages GPT-api to parse user instructions, yet it expects fully specified prompts (e.g., "extract baby crying from the audio" or "apply a low-pass filter to the wave crashing sound and add it back").
Some recent work~\cite{jia2024audioeditor, manor2024zero, xu2024prompt} adapts image-editing techniques (e.g., DDPM inversion, Null-text inversions, attention map manipulation) to monaural audio.
They require precise token-level swapping or deletion to complete the edit, struggle with high-level instructions.
Besides, they offer no support for stereo audio editing that are important in many applications~\cite{lan2024acoustic, xu2025mospa, liang2025binauralflow}.
To summarize, none of the existing work can interpret high-level user input to complete the audio editing.
Besides, existing frameworks remain confined to monaural outputs and are ill-suited for immersive spatial scenarios.

\header{Multimodal Large Language Model.}
Large language models (LLMs) are remarkable in natural language processing tasks. 
Provided with multimodal inputs, such as images and audio, multimodal LLMs (MLLMs)~\cite{alayrac2022flamingo, liu2023llava, li2023multimodalfoundationmodels, Qwen2.5-VL,kong2024audioflamingo, ghosh2025audioflamingo2} demonstrate exceptional performance across a wide range of downstream visual-language and audio-language tasks. 
In the vision domain, LLaVA~\cite{liu2023llava} enables LLMs to achieve general-purpose visual and language understanding by fine-tuning on a multimodal instruction-following dataset. 
In the audio domain, LTU~\cite{gong2023listen} and Audio Flamingo~\cite{kong2024audioflamingo, ghosh2025audioflamingo2} enhance LLMs with the ability to process non-speech sounds and non-verbal speech. 
With strong capabilities of MLLMs, researchers introduced them into the field of content generation \cite{wu2024videoinpaintingllm, wu2024visionllm, koh2023gill, fu2024mgie, huang2023smartedit}, grounding \cite{Lai_2024_CVPR, hao2024cvpr, Cheng_2024_CVPR}, world modeling \cite{wu2024ivideogpt, ge2024worldgpt, mai2024efficientmmworld}, and embodied AI \cite{driess2023palme, li2023manipllm, mu2023embodiedgpt}.
In image generation and editing, various works \cite{koh2023gill, fu2024mgie, huang2023smartedit} use VLM to guide diffusion models. 
However, in the audio domain, existing methods have not exploited the reasoning capabilities of audio language models. 

\header{Diffusion-based Image Editing.}
In contrast to text-to-image generation, image editing focuses on altering specific elements or attributes within an image while preserving the contents of the remaining image. 
Diffusion models have been widely used in image editing tasks \cite{couairon2022diffeditdiffusionbasedsemanticimage, hertz2022prompt2prompt, hui2024hq} by altering the inversion process, which produces a latent representation that can reconstruct the image through the generative process. 
SDEdit~\cite{meng2022sdedit} first adds noise to the source image, and then subsequently denoises the image through the SDE to produce the target image. 
P2P~\cite{hertz2022prompt2prompt} adjusts the cross-attention features according to the difference between the source and target captions to generate the target images. 
Based on this, IP2P~\cite{brooks2023instructpix2pix} finetuned a diffusion model on edit image triplets to enable image editing with simple natural language instructions.
Following works~\cite{geng2023instructdiffusion, hui2024hq} further scale up the dataset to support more capable and generalized models. 
Furthermore, \cite{fu2024mgie, huang2023smartedit} use vision-language models to guide diffusion models for image editing tasks. 

%% file: sections/03_method.tex
\vspace{-5pt}
\section{Method}
\vspace{-5pt}
In this section, we first define the task and notations.
We then introduce our proposed framework \name{} that combines an Audio Language Model for interpreting high-level editing instructions with a diffusion model for executing sequential audio edits.
Finally, we describe a scalable data generation pipeline powered by LLMs to support supervised training and evaluation.

\vspace{-5pt}
\subsection{Problem Definition and Notations}
\vspace{-5pt}
Let $a_0$ denote the original audio waveform, which contains multiple audio events (e.g., \emph{cat meowing}, \emph{rainfall}), as shown in Fig.~\ref{fig:teaser}. 
We define a \emph{high-level editing instruction} $\mathcal{P}$ as a natural language description of a desired transformation of the overall audio scene. 
Such instructions are typically declarative: they specify what the desired outcome should be (e.g., “\textit{make it sound like a quiet morning in a sunny forest}” or “\textit{transform this into an indoor library setting}”), but do not explicitly prescribe the individual operations.
Since $\mathcal{P}$ is a high-level instruction, it must be decomposed into a detailed sequence of atomic editing steps $\mathcal{S} = \{s_1, s_2, \ldots, s_n\}$, where each step $s_i$ either modifies an existing audio event in $a_0$ or introduces a new event needed to fulfill $\mathcal{P}$. 
We denote $a_i(i\!=\!1,2,...,n)$ as the intermediate audio after applying step $s_i$, where $a_0$ is the original input and $a_n$ is the final edited audio. 

Specifically, each step $s_i$ either modifies an existing audio event or introduces a new event required to satisfy $\mathcal{P}$.
The atomic editing operations considered in this work are:
\begin{itemize}[leftmargin=*, itemindent=0pt, itemsep=2pt, topsep=2pt, parsep=0pt]
\vspace{-2pt}
\item \texttt{Add:} Mix a new sound event into the scene (e.g., inject bird chirps).
\item \texttt{Remove:} Delete an existing sound event (e.g., remove car engine noise).
\item \texttt{Extract:} Isolate a particular sound event from the original audio while removing others.
\item \texttt{Turn volume up/down:} Adjust the volume of a specific event.
\item \texttt{Change direction:} Modify the spatial location of an event.
\vspace{-2pt}
\end{itemize}

The goal of the editing is to produce a target edited audio clip $a_n$ by applying the sequence of edits $\mathcal{S}$ to $a_0$.
Importantly, this editing formulation must preclude shortcut solutions that ignore the original input (e.g., re-generating an entirely new clip from scratch).
Instead, the edited audio $a_n$ must preserve all unedited content from $a_0$ while achieving the requested audio scene transformation. Please refer to Appendix \ref{app:atomic actions} for more details.

\vspace{-5pt}
\subsection{SmartDJ Framework}

\name{} consists of an Audio Language Model (ALM) and a Latent Diffusion Model (LDM).
We leverage the ALM as a planner to interpret high-level instructions and generate a sequence of 
\begin{wrapfigure}[11]{r}{0.4\textwidth} 
  \vspace{-14pt}                  
  \centering
  \includegraphics[width=0.4\textwidth]{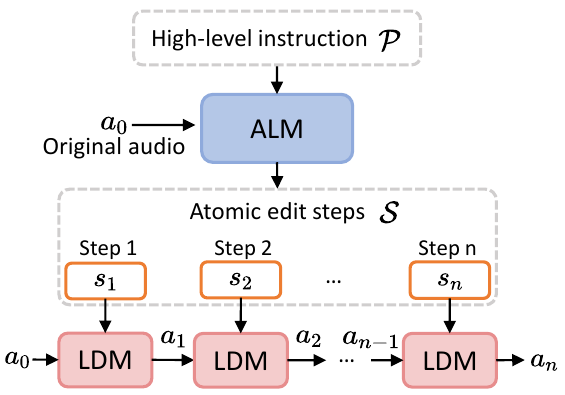}
  \vspace{-20pt}
  \caption{\name{} framework overview}
  \label{fig:overview}
\end{wrapfigure}
atomic editing steps. The LDM editor then executes these steps sequentially to transform the original audio.
As illustrated in Fig.~\ref{fig:overview}, the ALM takes the original audio $a_0$ and the high-level editing instruction $\mathcal{P}$ as input, and outputs a sequence of atomic editing steps $\mathcal{S} = \{s_1, s_2, ..., s_n\}$.  
These editing steps are then executed sequentially by the LDM, producing intermediate results $a_1, a_2, \ldots, a_n$, where $a_n$ is the final edited audio.  
The overall process is formulated as:
\begin{align}
    \{s_1,\!s_2,\!...,\!s_n\} \!&=\! \text{ALM}(a_0; \mathcal{P}) \\ 
    a_i \!&=\!\text{LDM}(a_{i-1};\!s_i), i\!=\!1,\!2,...,\!n
\end{align}

\subsection{Audio Language Model for Atomic Editing Steps Generation}
\begin{figure}[t]
    \vspace{-10pt}
    \centering
    \includegraphics[width=0.95\linewidth]{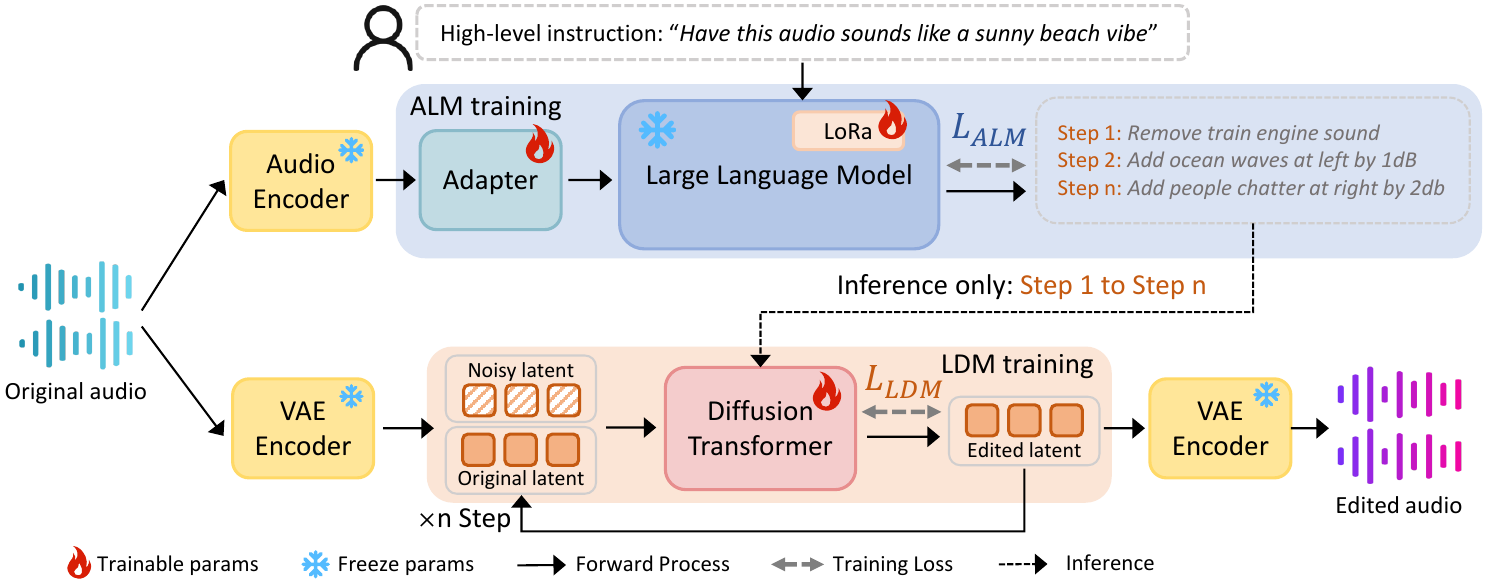}
    \vspace{-5pt}
    \caption{\textbf{\name{} framework}. Our method incorporates an ALM as an edit planner that understands both the original audio and the high-level instructions to produce atomic edit steps. These atomic steps are then fed into an LDM editor to edit the audio sequentially. The ALM and LDM modules are trained separately.}
    \vspace{-5pt}
    \label{fig:framework}
    \vspace{-8pt}
\end{figure}

\vspace{-5pt}
The Audio Language Model (ALM) takes as input the original audio clip and the high-level editing instruction and generates a sequence of atomic editing steps.
As shown at the top of Fig.~\ref{fig:framework}, we first encode $a_0$ using a pretrained audio encoder (i.e., CLAP~\cite{wu2023large}) to obtain an audio embedding $z_a$, which is injected into the ALM via adapter layers.
In parallel, the instruction $\mathcal{P}$ is tokenized and encoded as a sequence of embeddings $(p_1, p_2, \dots, p_k)$, which serve as the textual context for the ALM.

Our ALM is trained in an auto-regressive manner to generate the token sequence corresponding to the atomic editing steps $\mathcal{S}$, by minimizing the following objective:
\vspace{-5pt}
\begin{equation} 
\mathcal{L}_{\text{ALM}} = -\sum_{t=1}^l \log P_\theta(r_t'\!=\!r_t \mid z_a, r_{1:t-1}, p_{1:k}),
\end{equation}
where $r$ and $r'$ are the ground truth and predicted text tokens for the atomic editing steps $\mathcal{S}$, $l$ is the length of the tokens, 
and $\theta$ denotes the model parameters. 
To enable efficient fine-tuning, we freeze the parameters of the CLAP audio encoder, apply Low-Rank Adaptation (LoRA)~\cite{hu2022lora} to a small subset of the LLM layers\cite{ghosh2025audioflamingo2}, and fully fine-tune the adapter layers.

\header{Separate training.} 
We train the ALM and LDM as independent modules rather than end-to-end. 
This enables human-in-the-loop editing, where users can easily intervene at the level of generated natural-language-based atomic steps before LDM editor inference. 
Besides, it makes training and computation more efficient and promotes modularity. This design allows for different ALMs or LDMs to be swapped in or out with minimal re-training effort, making it both practical and extensible for diverse editing scenarios.

\vspace{-10pt}

\subsection{Sequential Stereo Audio Editing with Latent Diffusion Model}
\vspace{-5pt}
The Latent Diffusion Model (LDM) in our framework performs audio editing conditioned on the atomic editing steps $\mathcal{S}$.  
To support this, we adopt a latent diffusion architecture~\cite{hai2024ezaudio, rombach2022high} and extend it to enable editing of stereo audio with spatial effects.

\noindent\textbf{Stereo Audio VAE.}  
Given a stereo audio signal $a \in \mathbb{R}^{2 \times L}$, where $L$ is the number of time-domain samples at two audio channels (left and right).  
The audio Variational Autoencoder (VAE) encodes $a$ into a latent representation $\hat{a} \in \mathbb{R}^{C \times L'}$, where $C$ and $L'$ denote the number of latent channels and the temporal length of the latent sequence.  
Similar to DAC~\cite{kumar2023high} and Stable Audio Open~\cite{evans2024stable}, our audio VAE is based on a 1D-CNN autoencoder with a continuous VAE bottleneck and snake activation functions.  
The resulting latent $\hat{a}$ has a dimension of $C\!=\!128$ and length of $L'\!=\!L / 480$, resulting a compression ratio of 7.5$\times$.

\noindent\textbf{Latent Diffusion Model:}
Our diffusion model conditions on both the text description $s_i$ at the $i$-th editing step and the latent representation of the audio from the previous step, $\hat{a}_{i-1}$, to generate the updated latent $\hat{a}_i$.  
We use the FLAN-T5~\cite{chung2024scaling} text encoder $E_\text{text}()$ to convert $s_i$ to text embeddings.
At each editing step, we initialize a randomly noised latent $\hat{a}'_i \in \mathbb{R}^{C \times L'}$, which is concatenated with $\hat{a}_{i-1}$ to form the input $[\hat{a}_{i-1}; \hat{a}'_i] \in \mathbb{R}^{2C \times L'}$ to the diffusion model.  
The model is conditioned on $E_\text{text}(s_i)$ via cross-attention layers, and the diffusion timestep $t$ is incorporated through a modified AdaLN module~\cite{hai2024ezaudio} to reduce model parameters.
We implement a Diffusion Transformer (DiT) that learns to denoise the latent $\hat{a}'_i$ across multiple timesteps by predicting the added noise.  
Let $\epsilon$ denote the true added Gaussian noise, and let $\epsilon_\theta(\cdot)$ be the predicted noise output by the model.  
The training objective is to minimize the following denoising loss:
\begin{equation}
    \mathcal{L}_{\text{LDM}} = \mathbb{E}_{\epsilon\sim\mathcal{N}(0, I), t, s_i, \hat{a}_{i-1},\hat{a}'_{i}}\left \| \epsilon - \epsilon_\theta(t, E_\text{text}(s_i), [\hat{a}_{i-1}; \hat{a}'_{i}]) \right \|_2.
\end{equation}

During inference, we use DDIM sampling~\cite{song2021ddim} with classifier-free guidance (CFG), which has proven effective for text-guided generation and editing~\cite{ho2022classifier}.  
CFG steers the denoising process by interpolating between conditional and unconditional model predictions:
\begin{equation}
    \Tilde{\epsilon}_\theta =\omega\cdot\epsilon_\theta(t,  E_\text{text}(s_i), [\hat{a}_{i-1}; \hat{a}'_{i}]) + (1-\omega)\cdot\epsilon_\theta(t, \varnothing, [\hat{a}_{i-1}; \hat{a}'_{i}]),
\end{equation}
where $\omega$ is the guidance scale and $\varnothing$ denotes the text embedding of an empty string.

\vspace{-5pt}
\subsection{High-level Instruction Audio Editing Dataset Curation}
\vspace{-5pt}

\begin{figure}[t]
    \centering
    \includegraphics[width=0.9\linewidth]{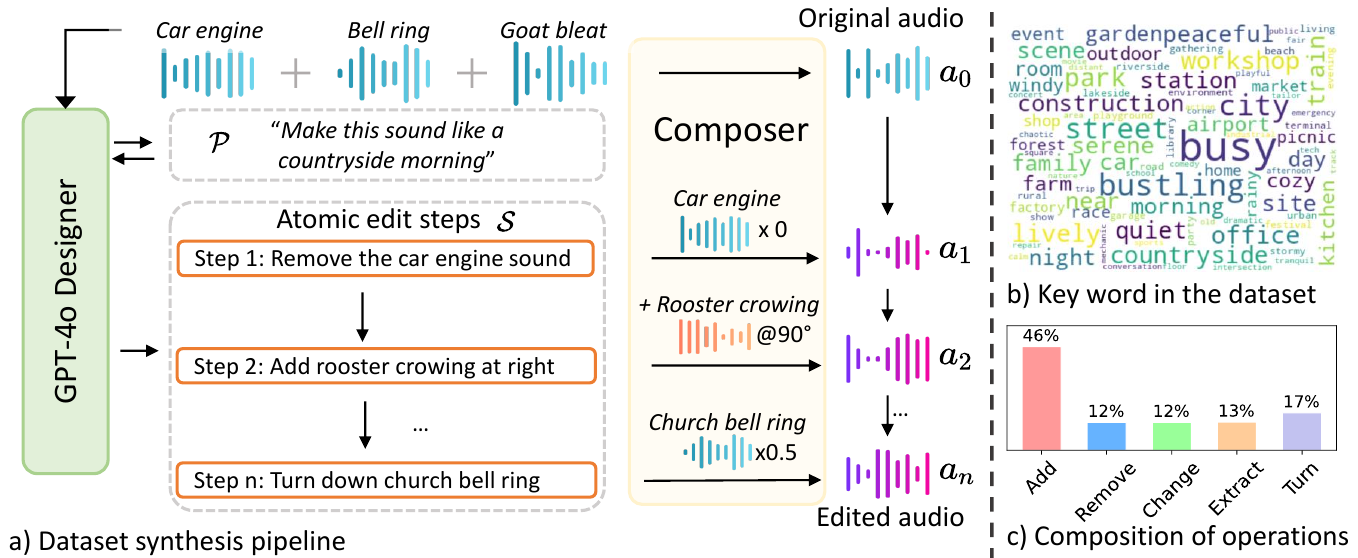}
    \vspace{-5pt}
    \caption{\textbf{Scalable data synthesis.} a) pipeline: we sample audio clips from databases with text labels and compose them into original audio $a_0$; These text labels are then fed into GPT-4o, which is prompted to design a high-level instruction $\mathcal{P}$ and generate corresponding atomic steps $\mathcal{S}$. We compose the target audio $a_1, a_2, ..., a_n$ following the atomic steps sequentially with rule-based composer. b) Key words in the high-level instruction. c) The proportion of each single-step edit operation in the dataset.}
    \label{fig:dataset}
    \vspace{-10pt}
\end{figure}

Since no public dataset features audio editing conditioned on high-level instructions, we develop a scalable data generation pipeline, as illustrated in Fig.~\ref{fig:dataset}a.
For each data point, we first randomly sample $K$ single-event audio clips from public datasets, each labeled with tags such as \{"\textit{car engine}", "\textit{bell ring}", "\textit{goat bleat}", ...\}.  
We feed these labels into GPT-4o and prompt it to act as a sound designer: it designs a high-level editing instruction $\mathcal{P}$ that transforms the original mix into a new audio scene (e.g., \textit{"Make this sound like a countryside morning"} or \textit{"Make it sound like a busy train station on a sunny afternoon"}). 
It then decomposes $\mathcal{P}$ into a sequence of atomic edits $\mathcal{S} = \{s_1, s_2, ..., s_n\}$, including both \texttt{add} operations and modifications to existing events (e.g., \texttt{remove}, \texttt{turn up/down}, \texttt{change sound direction}).

Once the sound designer has provided the sequential edit operations, an audio signal processing based composer takes over.
To generate edit audio pairs, the composer first synthesizes the initial audio $a_0$ by superposing the $K$ audio clips.
Spatial effects are rendered with direction-dependent phase and amplitude on two channels~\cite{lan2024acoustic}. 
For each atomic edit $s_i$, we proceed as follows:  
\begin{itemize}[leftmargin=*, itemindent=0pt, itemsep=2pt, topsep=2pt, parsep=0pt]
\vspace{-5pt}
    \item If $s_i$ modifies an existing event, we update its volume level or sound direction.
    \item If $s_i$ is an \textit{add} operation, we retrieve a new clip from the database with a matching label.
\vspace{-5pt}
\end{itemize}
Since each sampled audio event is \textit{independently editable}, an edit step $s_i$ that modifies an existing event can be converted into event-level parameter adjustments, without altering any other sources in the mixture $a_{i-1}$.  
For example, to simulate \textit{"remove the car engine sound"} (Fig.~\ref{fig:dataset}), we set the car engine’s volume to zero to generate $a_1$.
To simulate \textit{"add rooster crowing at right"}, we retrieve a rooster clip, apply the specified spatial effect, and superpose it on $a_1$ to obtain $a_2$.  
To simulate \textit{"turn down the dog bark"}, we reduce the volume of the corresponding clip.
This allows us to generate a complete editing trajectory $a_1, a_2, ..., a_n$ by progressively updating event-level parameters and re-composing the audio scene.  
Our resulting dataset contains high-level editing instructions, atomic edit sequences, and the audio for the full editing trajectory. More details available in Appx. \ref{app:complex editing dataset}. 

%% file: sections/04_experiments.tex
\vspace{-5pt}
\section{Experiment}
\vspace{-10pt}
\subsection{Setup}
\vspace{-10pt}

\noindent\textbf{Dataset}. We use a combination of datasets including AudioCaps~\cite{kim2019audiocaps}, VGGSound~\cite{chen2020vggsound}, FSD50k~\cite{fonseca2021fsd50k}, ESC50~\cite{piczak2015esc}, and WavCaps~\cite{mei2024wavcaps}. 
We adopt a series of dataset cleaning pipelines following previous work~\cite{hai2024ezaudio, sun2024both, wang2023audit} by filtering events with noisy data labels or low clap scores. 
Each audio is trimmed or padded into 10 seconds with a sampling rate of 24~K.
We sample 2-5 audio events and use GPT-4o to create 50k training pairs and 1k evaluation pairs of high-level audio editing data to train audio language model and evaluate the whole editing pipeline. 
We present the keyword in the high-level instructions in Fig.~\ref{fig:dataset}b.
We also expand the size of single-step editing data pairs ($s_t, a_{t-1}, a_{t}$) to 0.5M, where each step is a single atomic operation to train our LDM audio editor.
Fig.~\ref{fig:dataset}c shows the proportion of each operation.
Please find more details in the Appx. \ref{app:dataset preparation}.

\header{Metrics}. To evaluate edit quality and diversity, we use common metrics in audio generation and editing~\cite{liu2023audioldm, wang2023audit}, including Fréchet Distance (FD), Kullback-Leibler divergence (KL), Fréchet Audio Distance (FAD), Inception Score (IS), and Log-Spectral Distance (LSD). We use CLAP score to measure the semantic similarities between the edited audio and the text prompt.
For spatial audio, we calculate GCC MSE (GCC) based on Generalized Cross-Correlation with Phase Transform (GCC-PHAT), and use StereoCRW~\cite{chen2022sound} to produce stereo audio features to evaluate CRW MSE (CRW) and Fréchet Stereo Audio Distance (FSAD).

\header{Baselines}. 
To evaluate the high-level instruction based audio editing task, we first train an end-to-end version of Audit~\cite{wang2023audit} that directly predicts the final-step edited audio $a_n$ conditioned on the original audio $a_o$ and high-level instruction $\mathcal{P}$ in one step without ALM. 
We extend the mono-channel Audit to our binaural setting, where we stack the left and right channels of the mel-spectrograms as the model inputs.
We also evaluate various zero-shot and training-required editing methods based on the ALM's outputs to perform multi-step sequential editing. 
The zero-shot methods include SDEdit~\cite{meng2022sdedit}, DDIM Inversion~\cite{mokady2022nulltextinversion}, ZETA~\cite{manor2024zero}, and AudioEditor~\cite{jia2024audioeditor}.
For a fair comparison, we replace these methods' generation backbone with the current SOTA methods. 
In AudioEditor, we replace Affusion with BEWO~\cite{sun2024both} to support binaural editing. 
In SDEdit, DDIM Inversion, and ZETA, we use Stable-Audio-Open~\cite{evans2024stable} as the backbone.
For sequential editing with a trainable editor, we also use an Audit baseline trained on single-step audio editing. 
In addition, we evaluate the same set of baseline methods on single-step audio editing tasks. More details available in Appx. \ref{app:baseline implementation}.

\header{Implementation details.} 
\name{} ALM is initialized from Audio Flamingo 2~\cite{ghosh2025audio} with 3B parameters. 
During training, we freeze the AF-CLAP module and fine-tune the adapter layers and LLM with a LoRA module for 20 epochs with a batch size of 24.
For LDM, it uses velocity prediction with Zero-SNR, and CFG rescaling technique~\cite{lin2024common} to adjust the magnitude of the predicted velocity and avoid over-exposure.
It is trained on single-step editing data with a batch size of 256 of 500k iterations.
10\% text is replaced with empty strings to enable unconditional modeling. 
The learning rates for the ALM and LDM training are 1e-5 and 5e-5 with AdamW.
All experiments are conducted with four NVIDIA L40S GPUs. More details available in Appx. \ref{app:smartdj implementation}.

\begin{figure}[!t]
    \centering
    \vspace{-5pt}
    \includegraphics[width=0.85\linewidth]{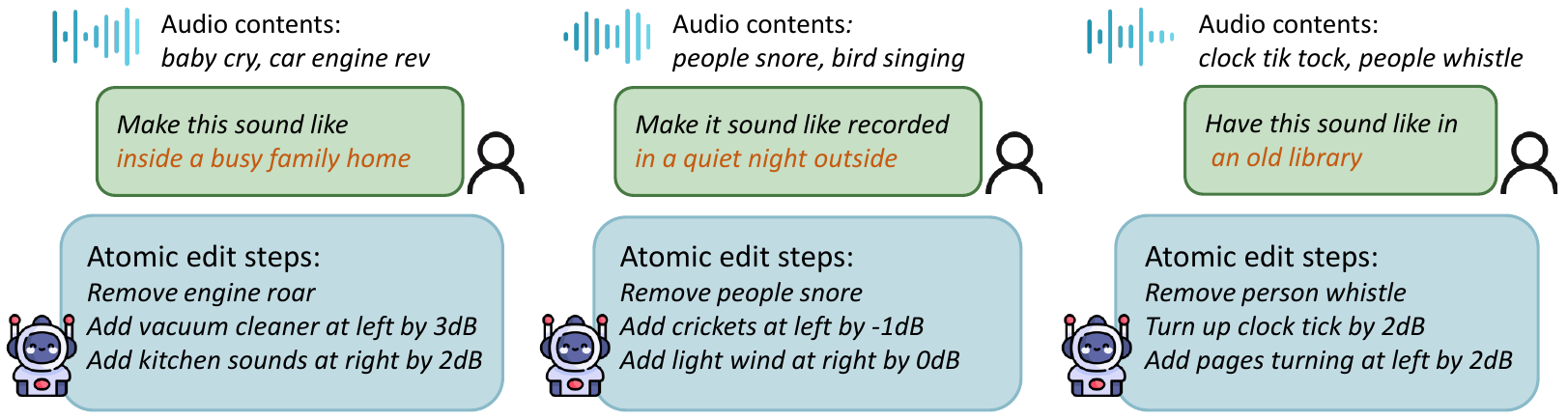}
    \vspace{-10pt}
    \caption{Examples of ALM's output detailed steps. Our ALM module identifies events in the original audio clips and reasons on the given high-level instruction to produce aligned editing steps.}
    \label{fig:llm_examples}
    \vspace{-13pt}
\end{figure}

\vspace{-5pt}
\subsection{Results}
\vspace{-5pt}
\header{High-level instruction audio editing task.}
We first show inference examples from our ALM module in Fig.~\ref{fig:llm_examples}.
The ALM-generated atomic editing steps accurately align both the original audio contents and the high-level editing instructions.
For example, it correctly removes audio event \textit{engine roar} when transferring audio scene into a \textit{busy family home} vibe.
It also removes \textit{people whistle} and adds \textit{pages turning} to enhance the immersion of being in an \textit{old library}. More examples in Appx. \ref{app:alm_inference_results}.

\begin{table*}[h]
\centering
\vspace{-5pt}
\begin{adjustbox}{max width=.9\textwidth}
\small
\begin{tabular}{@{}cccccccccc@{}} 
\toprule
Framework & Method & Training & Speed & FD $\downarrow$ & FAD  $\downarrow$& KL$\downarrow$ & LSD$\downarrow$ & IS $\uparrow$& CLAP$\uparrow$  \\ 
\midrule
\multirow{1}{*}{w/o ALM} & Audit & \cmark & 2.07s & 39.6 & 10.07 & 3.13  & 1.96   & 3.29 & 0.125    \\\midrule
\multirow{6}{*}{w/ ALM} & SDEdit & \xmark & 301s (74.6s) & \underline{27.3} & \underline{3.73} &3.26 &2.25 &6.66 & 0.188 \\
 &DDIM & \xmark & 331s (82.1s) & 34.3 &9.49 &4.07 &2.23 &3.97 & 0.076 \\
 &ZETA & \xmark & 356s (88.2s) & 28.8 & 3.75 &2.93 &2.24 &7.72 & \underline{0.224} \\ 
 &AE & \xmark & 406s (101s) & 27.6 &5.02 &3.22 & 2.11 & \textbf{8.91} & 0.211 \\
 &Audit & \cmark & 11.6s (2.07s) & 29.4 &5.71 & \textbf{2.81} & \underline{1.51} &3.95 & 0.197 \\
 &SmartDJ (Ours) & \cmark & 13.1s (2.40s) & \textbf{14.7} & \textbf{1.53} & \underline{2.85} & \textbf{1.42} & \underline{8.36} & \textbf{0.238} \\
\bottomrule
\end{tabular}
\end{adjustbox}
   \vspace{-5pt}
\caption{Quantitative results of the whole pipeline from high-level instructions to audio editing. Speed in () is the time for a single-step edit. AE denotes AudioEditor; DDIM denotes DDIM Inversion.}
\label{tab:whole_system}
\vspace{-8pt}
\end{table*}

We present the results of the high-level instruction audio editing task in Tab.~\ref{tab:whole_system}. 
The first row is the end-to-end Audit baseline, which iss directly trained on high-level instructions and the final target audio, showing the worst performance overall. 
Since no prior method can interpret the high-level instructions, we use the same set of ALM-generated atomic steps to guide all audio editing models in the multi-step evaluation, including the baselines and our method.
We compare the edited audios from each method with 1k reference audios.
Our method achieves the lowest metric in FD, FAD, LSD, and comparably low in KL, indicating the smallest discrepancy from the reference audios.
It also delivers a high IS metric, showing strong audio quality and diversity.
In addition, the highest CLAP score demonstrates the best semantic alignment between the edited audio and instruction.

\header{Inference time.} We also provide inference speed analysis in Tab.~\ref{tab:whole_system}. 
The inference speed reported outside the brackets denote the total time to complete the entire high-level instruction edit, while the values inside indicate the per-round inference time of LDM. 
On average, \name{}’s ALM requires 4.8s to generate one set of atomic editing instructions. 
With multi-round reasoning, \name{} completes a full edit in 13.1s, which is significantly faster than training-free baselines. 
Our approach is slower compared with end-to-end Audit (2.07s) or Audit with ALM (11.6s), but this trade-off yields substantially better editing quality and alignment with target instructions. 

\begin{figure}[!t]
    \centering
    \includegraphics[width=0.95\linewidth]{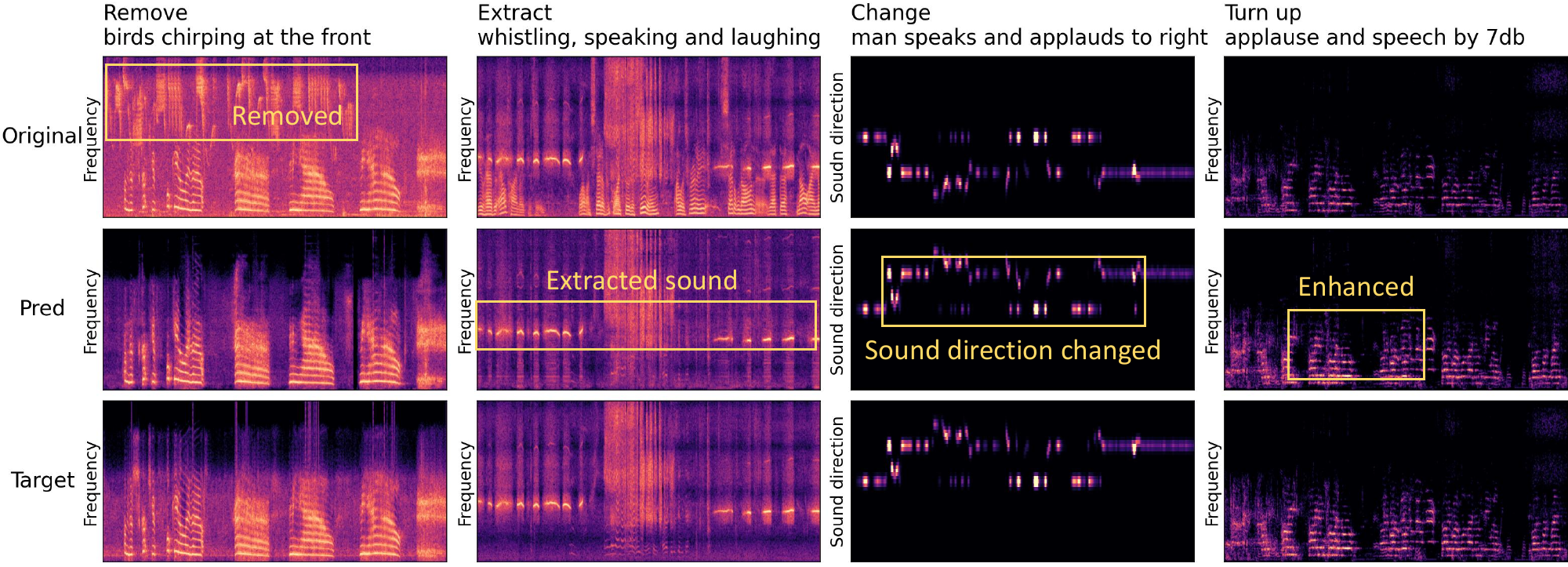}
    \vspace{-5pt}
    \caption{Examples on \texttt{Remove}, \texttt{Extract}, \texttt{Change}, \texttt{Turn up} operations. \name{} edited audio are closely aligned with the ground truth. Yellow boxes highlight edited sound components.}
    \label{fig:edit_examples}
    \vspace{-8pt}
\end{figure}

\begin{table}[t]
\vspace{-2pt}
    \centering
    \begin{subtable}[t]{\textwidth} 
    \centering
    \begin{adjustbox}{max width=\linewidth}
    \begin{tabular}{@{}cccccccccccccccc@{}} 
    \toprule
    \multirow{2.5}{*}{Method} & \multicolumn{8}{c}{\texttt{Add}} & \multicolumn{7}{c}{\texttt{Remove/Extract}} \\ 
    \cmidrule(lr){2-9}
    \cmidrule(lr){10-16}

     & FD $\downarrow$ & FAD $\downarrow$ & KL $\downarrow$ & LSD $\downarrow$ & IS $\uparrow$ & GCC $\downarrow$ & CRW $\downarrow$ & FSAD $\downarrow$ & FD$\downarrow$ & FAD$\downarrow$ & KL$\downarrow$ & LSD$\downarrow$ & GCC$\downarrow$ & CRW$\downarrow$ & FSAD$\downarrow$   \\ 
    \midrule
SDEdit & 33.0 & 3.92 & 2.59 & 2.01 & 4.71 & 143.3 & 131.4 & 0.42 & 46.6 & 4.64 & 2.38 & 1.89 & \underline{159.5} & 157.9 & 0.45 \\
DDIM & 36.9 & 6.28 & 2.64 & 2.02 & 4.57 & \underline{131.2} & \underline{116.1} & \underline{0.11} & 53.9 & 5.69 & 2.75 & 1.85 & 159.6 & \underline{138.5} & \underline{0.29} \\
ZETA & 37.6 & \underline{3.64} & 2.46 & 1.71 & 5.04 & 143.3 & 127.6 & 0.52 & \underline{40.6} & 3.78 & \underline{1.92} & 1.99 & 161.7 & 155.9 & 0.43 \\
AE & \underline{30.5} & 3.66 & 2.13 & 1.84 & \underline{5.51} & 133.5 & 145.8 & 0.55 & 47.1 & \underline{3.65} & 2.43 & 2.01 & 164.0 & 165.6 & 0.58 \\
Audit & 36.5 & 4.49 & \underline{1.95} & \underline{1.42} & 4.37 & 145.7 & 136.3 & 0.31 & 54.5 & 7.56 & 2.12 & \textbf{1.65} & 189.2 & 204.5 & 1.07 \\
SmartDJ & \textbf{22.7} & \textbf{1.82} & \textbf{1.39} & \textbf{1.35} & \textbf{5.96} & \textbf{76.5} & \textbf{41.3} & \textbf{0.03} & \textbf{26.0} & \textbf{2.57} & \textbf{1.03} & \underline{1.71} & \textbf{15.9} & \textbf{5.5} & \textbf{0.02} \\\bottomrule
     \end{tabular}
     \end{adjustbox}
     \vspace{-2pt}
    \caption{\begin{small}Audio editing operation \texttt{add} and \texttt{remove/extract}\end{small}}
    \label{tab:detailed_editing_1}
    \end{subtable}
    \vspace{-2pt}

    \begin{subtable}[htbp!]{\textwidth} 
    \centering
    \begin{adjustbox}{max width=0.9\linewidth}
    \begin{tabular}{@{}ccccccccccccccc@{}} 
    \toprule
    \multirow{2.5}{*}{Method} & \multicolumn{7}{c}{\texttt{Turn Up/Down}} & \multicolumn{7}{c}{\texttt{Change}} \\ 
    \cmidrule(lr){2-8}
    \cmidrule(lr){9-15}
      & FD$\downarrow$ & FAD$\downarrow$ & KL$\downarrow$ & LSD$\downarrow$ & GCC$\downarrow$ & CRW$\downarrow$ & FSAD$\downarrow$ & FD$\downarrow$ & FAD$\downarrow$ & KL$\downarrow$& LSD$\downarrow$& GCC$\downarrow$ & CRW$\downarrow$ & FSAD$\downarrow$  \\ 
    \midrule
    Audit & 47.1 & 5.6 & 1.51 & 1.04 & 136.8 & 139.0 & 0.89 & 42.5 & 4.9  & 1.34 & 1.02 & 170.3 & 163.5 & 0.99\\
    SmartDJ & \textbf{11.8}  & \textbf{1.0} & \textbf{0.27} & \textbf{1.01} & \textbf{23.3} & \textbf{2.86} & \textbf{0.01} & \textbf{13.0}  & \textbf{0.88} & \textbf{0.33} & \textbf{1.00} & \textbf{59.6} & \textbf{36.6} & \textbf{0.02} \\
    \bottomrule
     \end{tabular}
     \end{adjustbox}
         \vspace{-2pt}

    \caption{\begin{small}Audio editing operation \texttt{turn up/down}, \texttt{change sound direction}\end{small}}
    \label{tab:detailed_editing_3}
    \end{subtable}
    \vspace{-10pt}
    \caption{Quantitative results on all individual audio editing operations. }
    \label{tab:detailed_editing_all}
\vspace{-15pt}
\end{table}

\header{Single-step audio editing.}
We present the results of single-step audio editing operations.
Tab.~\ref{tab:detailed_editing_1} shows results on \texttt{add}.
Our method consistently outperforms baseline methods in the edited audio, including better similarities to the ideal target audio (lowest FD, FAD and KL), and higher quality and diversity shown by the highest IS. 
Furthermore, stronger spatial metrics also indicate that the stereo audio characteristics are preserved better by \name{}. 

Tab.~\ref{tab:detailed_editing_1} also shows the performance on \texttt{remove} and \texttt{extract} tasks. 
The results clearly indicate that our method delivers the best performance in aligning with the ground truth edited audio across both operations.
We show the results on \texttt{turn up/down} and \texttt{change sound direction} tasks in Tab.~\ref{tab:detailed_editing_3}.  
Our method again shows stronger performance over Audit, which demonstrates \name{} has better fine-grained manipulation in audio event properties.
This is because Audit's VAE operates on mel-spectrograms and discards phase information, which is critical for spatial cues.
\name{} employs a diffusion transformer, which provides stronger long-range temporal modeling and richer cross-attention conditioning.
These architectural upgrades allow edits to be both semantically precise and spatially coherent.
Some examples of spectrogram visualization are shown in Fig.~\ref{fig:edit_examples}. More qualitative comparisons can be found in Appx.~\ref{app:edit_examples}.

\begin{figure}[t]
  \centering
  \begin{minipage}[t]{0.63\linewidth}
    \centering
    \vspace{-5pt}
    \includegraphics[width=\linewidth]{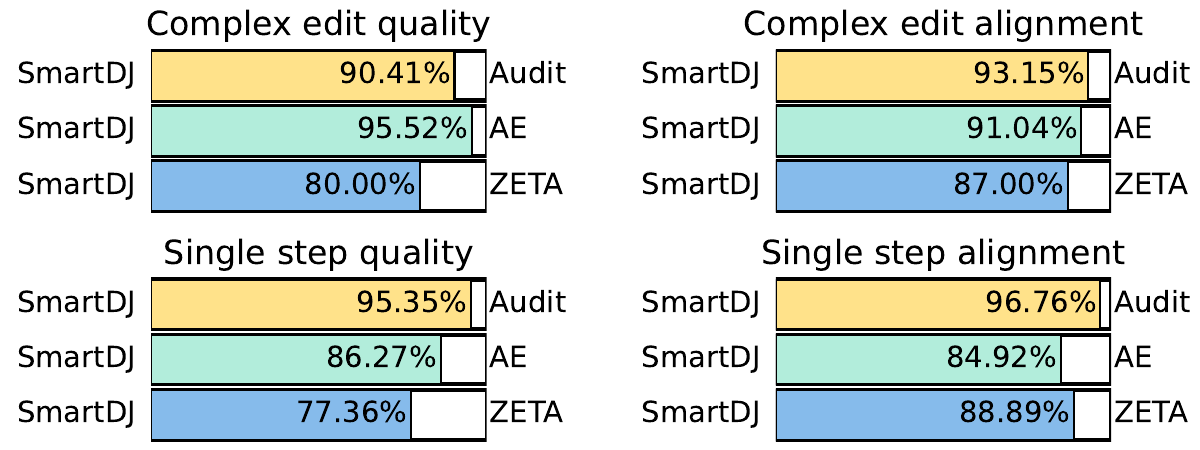}
    \vspace{-15pt}
    \caption{User study results. In both audio quality and text/audio alignment, \name{} is consistently preferred over baselines in the high-level instruction editing task and single-step tasks.}
    \label{fig:user_study}
  \end{minipage}
  \hfill
  \begin{minipage}[t]{0.33\linewidth}
    \centering
    \vspace{-5pt}
    \includegraphics[width=\linewidth]{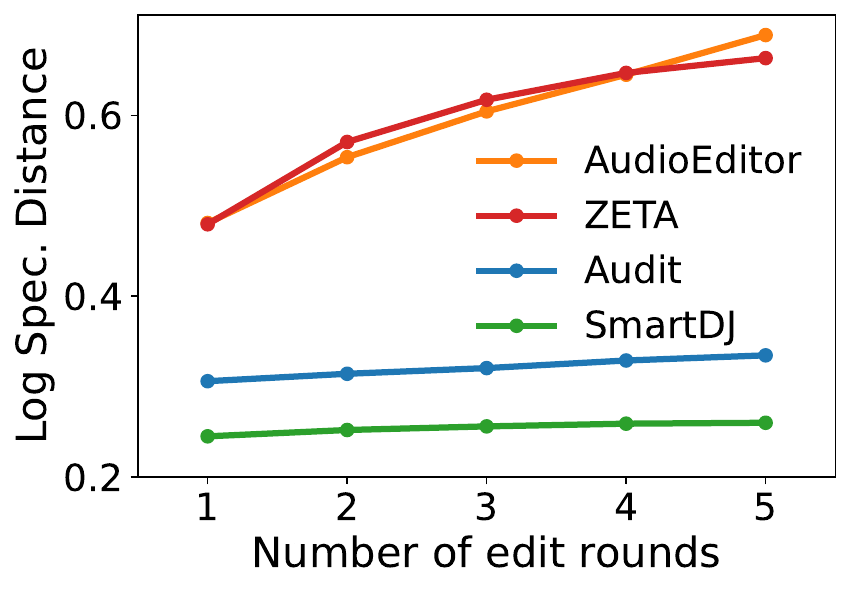}
    \vspace{-15pt}
    \caption{Similarity with original audio after multiple editing rounds.}
    \label{fig:multi_round_audio_quality}
  \end{minipage}
  \vspace{-5pt}
\end{figure}

\header{Human evaluations.} 
We evaluate the subjective preference via a user study.
We provide users with data pairs consisting of the original audio, the editing prompt, \name{} edited result, and edited results from a random competing baseline.
We ask the user to select the one that has higher audio quality, has better alignment with the text or spatial instructions, and aligns with the original audio.
We conducted extensive evaluations with 19 participants and 20 (10 high-level audio editing, 10 single-step editing) data pairs per participant. 
We separate the evaluation for high-level instruction audio editing and single-step editing in Fig.~\ref{fig:user_study}.
In both tasks, \name{} is much preferred over all competing methods.
Our method delivers the best audio quality and the best alignment with both the high-level instruction and the single-step instruction from user perception. More details in Appx.~\ref{app:human_subject_results}.

\vspace{-10pt}
\subsection{Ablation studies}
\vspace{-10pt}

\textbf{Audio quality over multi-round editing.} Since the high-level instruction audio editing task involves a sequence
of editing, unchanged content must remain intact after multiple steps. 
To evaluate this, we design a "round-trip" edit experiment: we perform the operations "add the sound of $\mathcal{A}$" and "remove the sound of $\mathcal{A}$" on an audio clip for five rounds, where $\mathcal{A}$ is a pseudo audio label.
For an ideal audio editor, this sequence of round-trip operation should exactly reconstruct the original audio.
We measure the LSD between each round's edited output with the original audio clip (Fig.~\ref{fig:multi_round_audio_quality}).
\name{} consistently achieves the lowest LSD, indicating the smallest drift from the original content. This demonstrates that our method preserves the unedited audio content under repeated editing actions.

\begin{table}[t] 
  \centering
  \begin{minipage}[t]{0.7\textwidth}   % LEFT: ablation table
    \centering
    \small
    \renewcommand{\arraystretch}{1.0}
    \resizebox{\linewidth}{!}{%
    \begin{tabular}{@{}cccccccc@{}}
      \toprule
      \textbf{Study Objectives} & \textbf{Variation} & FD$\downarrow$ & FAD$\downarrow$ & KL$\downarrow$ & LSD$\downarrow$ & IS$\uparrow$ & CLAP$\uparrow$ \\
      \midrule
      \multirow{2}{*}{ALM module} & \quad w/o ALM & 23.6 & 3.14 & 2.91 & 1.84 & 4.63 & 0.137 \\
      & \quad w/ ALM  & \textbf{14.7} & \textbf{1.53} & \textbf{2.85} & \textbf{1.42} & \textbf{8.36} & \textbf{0.238} \\
      \midrule
      \multirow{3}{*}{Editing order} & \quad \texttt{Add} $\to$ \texttt{Modify} $\to$ \texttt{Remove}   & 14.9 & 1.59 & 2.88 & 1.48 & 8.09 & 0.234 \\
        & \quad Random order & \textbf{14.7} & 1.56 & 2.88 & 1.47 & 8.16 & 0.237 \\
        & \quad \texttt{Remove} $\to$ \texttt{Modify} $\to$ \texttt{Add}  & \textbf{14.7} & \textbf{1.53} & \textbf{2.85} & \textbf{1.42} & \textbf{8.36} & \textbf{0.238} \\
      \midrule
      \multirow{2}{*}{\texttt{Extract} operation} & AudioSep~\cite{liu2024separate} & 27.1 & 2.86 & 0.88 & \textbf{1.63} & N/A & N/A \\
        & \name{} & \textbf{25.7} & \textbf{2.55} & \textbf{0.78} & 1.71 & N/A & N/A \\
      \bottomrule
    \end{tabular}}
    \vspace{-5pt}
    \caption{Model ablations.}
    \label{tab:ablation}
  \end{minipage}
  \vspace{-15pt}
\end{table}

\header{Effectiveness of ALM.} 
We conduct an ablation by removing the ALM module and training a variant of the LDM end-to-end.
As shown in Tab.~\ref{tab:ablation}, \name{} performs significantly worse without ALM, showing the importance of ALM’s intermediate reasoning capabilities.
ALM enables the model to produce semantically coherent edits aligned with the high-level instruction and the original audio.

\header{Editing order.}
We adopt a simple ordering strategy of remove $\rightarrow$ modify (volume/direction) $\rightarrow$ add, which intuitively avoids removing newly inserted content. 
To test the impact of ordering, we also experiment with two alternatives: randomized and reversed order. 
As shown in Tab.~\ref{tab:ablation}, both alternatives result in only marginal degradation compared to our default ordering. 
This indicates that the editing order has minimal influence on the final results, suggesting that ALM-generated instructions rarely contain conflicting operations.

\header{Compare with sound separation model.}
We compare \name{} with a recent sound separation model AudioSep~\cite{liu2024separate} on the \texttt{Extract} operation.
Ours achieves comparable or slightly better performance (Tab.~\ref{tab:ablation}).
This suggests that while our model is designed for general-purpose audio editing, it is competitive on target sound separation tasks.

%% file: sections/05_discussion.tex
\vspace{-10pt}
\section{Discussion}
\vspace{-15pt}

\header{Conclusion.} We presented \name{}, the first framework for high-level instruction guided stereo audio editing that utilizes the reasoning capability of audio language models and strong editing capabilities of the latent diffusion model. 
Our approach produces atomic editing steps and executes them sequentially to achieve perceptually realistic stereo audio transformations. 
Extensive evaluations on both subjective audio metrics and human perceptual studies demonstrate that \name{} outperforms prior methods, and preserves spatial fidelity in complex scenes. 

\header{Limitations.} Supporting a new task-specific editing operation on the LDM requires retraining the diffusion model. 
However, these edits can usually be achieved through a combination of our proposed atomic steps.
Besides, a future direction is to implement an end-to-end joint training strategy of ALM and LDM that combines reasoning with audio editing.

\section{Ethics statement.}
Our work focuses on audio editing for research and creative applications such as immersive media, conferencing, and sound design. The dataset is generated from publicly available sound event libraries and synthetic mixing, without personal or sensitive recordings. 
We encourage responsible use aligned with academic and creative purposes.

\section{Reproducibility statement}
We provide detailed descriptions of the model architectures, training objectives, and data generation pipeline in the main paper and appendix. Hyperparameters, training configurations, and dataset construction details are included to ensure reproducibility. Code, pretrained models, and the synthesized dataset will be released upon acceptance to facilitate replication of our results.

%% file: sections/06_appendix.tex
% \newpage 
\appendix
\section{Dataset curation process}

\subsection{Dataset preparation}
\label{app:dataset preparation}
We construct our training corpus by merging several publicly-available audio dataset, including VGG-Sound, AudioCaps, WavCaps, ESC-50, and FSD50K.
Since some of these sets provide audio captions rather than discrete audio label, we first convert every caption to audio labels with GPT-4o-mini API.
We only retain single-labeled audio clip and any clip whose caption maps to multiple events is discarded.
A CLAP model scores the semantic correspondence between the audio and its new label.
Samples with a similarity score below 0.3 are filtered out.
The remaining clips from all sources are finally mixed into one large dataset that we use for subsequent data curation.

\subsection{Atomic edit actions}
\label{app:atomic actions}
We explain the details on creating the single-step atomic edit data pairs.
For this single-step audio editing, we have an original audio $a_{i-1}$, a single atomic edit operation $s_i$.
Base on atomic edit operation $s_i$, we can generate the edited audio $a_i$.

\header{\textit{Add.}} Assume the original audio is a mix of audio content \(A\!+\! B\!+\!C\). To \textit{add} a new content into this original audio, we sample a new audio content D from the database and mix the D with the original audio \(\!A\!+\!B\!+\!C\!+\!D\).
The atomic template is \textit{"Add the sound of \{dog barking\} at \{right\} with \{3\} db"}. 
The contents inside the \{\} can be changed to other sound events or sound attributions.
We support various sound direction (left, front and right) and dynamic volume adjustment.

\header{\textit{Remove.}} Given an original mix \(A\!+\!B\!+\!C\), let \(B\) be the undesired source.
We suppress \(B\) so the output becomes \(A\!+\!C\).
THe atomic template is Atomic template:  
\textit{"Remove the sound of \{bird chirping\} \{at right\}"}.
The directional phrase in braces is optional.
If there are similar audio contents in the same clip, the spatial features enables to manipulate it precisely.

\header{\textit{Extract.}} Starting from the same mix \(A\!+\!B\!+\!C\), we isolate one target source \(A\) and mute everything else, yielding only \(A\).
The atomic template is \textit{"Extract the sound of \{speaking\} \{at the right\}"}. The direction is optional.

\header{\textit{Turn up/down}} To change loudness of a specific source \(B\), we scale it by \(\alpha\), where \(\alpha \!>\!1\) is for “up” and \(0\!<\!\alpha\!<\!1\) is for “down". The resulted audio clip is \(A + \alpha B + C\).
The atomic template is \textit{"Turn \{up / down\} the sound of \{engine rev\} by \{2\} dB"}.
We also support a dynamic range of volume adjustment.

\header{\textit{Change sound direction.}}
We alter only the spatial cues of a specific source \(C\), producing an edited
version \(C'\) while leaving the other tracks untouched:
\(A\!+\!B\!+\!C'\).
The atomic template used is \textit{"Change the sound of \{baby crying\} \{from front\} to \{right\}"}.
The “from” clause could also be omitted.

\subsection{Complex audio editing dataset curation}
\label{app:complex editing dataset}
In the dataset curation process, we first sample 2-5 audio labels in the database, with LLM randomly assigned volume and sound directions.
We then call the GPT-4o batch API with sound sources and attributions.
For each API call, we provide 15 sets of sound sources.
Through API call, each set of sound sources will return a single data pair containing high-level editing instruction and the corresponding atomic editing steps.
We follow these atomic editing steps to manually generate the step-by-step target edited audio with the rules in \ref{app:atomic actions}.
We generate 50K data pairs for complex audio editing for training.
We generate 1K data pairs from AudioCaps test/validation set for evaluation.

Starting from these 50K complex audio editing pairs, we collect the generated corresponding step-by-step atomic actions and produce about 200K single step audio editing pairs ($s_i, a_{i-1}, a_i$).
We further scale up the this dataset size to 500K to train the LDM audio editor and we also generate another 1K extra audio data pairs to evaluate the performance of single step audio editing.
This scaled-up single step audio editing dataset keeps some instructions with audio captions, improving the LDM's robustness to different audio contents in the atomic edit instructions.

We provide the details of our \textit{Base Prompt} for dataset curation as follow:
\begin{gptprompt}
You are an expert in spatial audio editing and sound design. 

Your task is to generate complex audio editing instructions based on a given list of sound sources (labels).
The sound sources will be provided as a list of full sentences (as strings), not character lists. Treat each sentence as a single atomic sound unit. Do not tokenize or split the sound sources into characters.

For example, if you are given: "a baby crying and a man talking; a bird is chirping; dog barking". 
You should consider "a baby crying and a man talking" as a complete sound source. "a bird is chirping" is another complete sound source and "dog barking" is another sound source.
You then generate the step-by-step audio editing instructions based on the given complex instructions.

Task: you need to first brainstorm a complex audio editing instruction
\begin{itemize}[leftmargin=1.5em, itemsep=1pt, label=-, topsep=0pt]
  \item Imagine a realistic and creative soundscape editing for the given audios.
  \item You are not limited to the provided audio contents for the editing instruction. And the complex editing instructions should be brief.
  \item The complex editing instructions could be a soundscape transformation. For example:
    \begin{itemize}[label=- -, itemsep=0.5pt, leftmargin=1.5em]
      \item Make this sound like it was recorded in a bookstore
      \item Make this sound like a busy coffee shop
      \item Make this sound like a train station
      \item Make this sound like a forest at night
      \item Make this sound like a beach
      \item Make this sound like a sunny day
      \item Craft this sound to feel like a park
      \item Make this audio sound like a quiet farm
      \item Make this audio sound like a firework show
    \end{itemize}
  \item Your generated complex instructions can be broader than these provided examples. Use your imaginations!
  \item But remember to keep them brief, the complex editing instructions ideally should not contain the actual sound sources.
  \item Then, based on the given sound event, give me the detailed editing instructions.   
\end{itemize}

You then generate detailed editing instructions based on the complex audio editing instruction.
\begin{itemize}[leftmargin=1.5em, itemsep=1pt, label=-, topsep=0pt]
  \item Ensure the editing makes sense (e.g., no waves in a desert, no sheep indoors, rustling leaves in the forest, seas waves in the beach, no raining in the sunny day).
  \item You are encouraged to remove the original audio contents, but you are required to maintain at least one sound source, not removing all of them.
  \item  Do not split or partially reference a sound source when applying operations.
  \item Use a combination of simple operations (but NOT necessarily all of them). For example:
   \begin{itemize}[label=- -, itemsep=0.5pt, leftmargin=1.5em]
      \item Add (e.g., thunderstorm, cat meowing) (for the add operation, you should add up to two additional sources that best align with the complex editing instructions)
      \item Remove (For remove, you should remove at least one sound sources (up to two sound sources), but you must keep at least one sound source!)
      \item Turn up/down (e.g., turn up/turn down the sound of xxx by xxx db (between 0 and 6db))
      \item Change sound direction.
    \end{itemize}
  \item Each "target" in remove/turn up/turn down/change must exactly match one of the provided "sound sources"
  \item For the "add" operation:
   \begin{itemize}[label=- -, itemsep=0.5pt, leftmargin=1.5em]
      \item The "target" must clearly describe the new sound being added (e.g., "crowd chatter", "rain", "footsteps on gravel").
      \item The "effect" should specify the volume and direction (e.g., "at front by 4dB").
      \item Do NOT use "none", "null", or placeholder values as the target. The target must always be a descriptive label of the added sound.
      \item  The added sound should not duplicate any original sound source.
    \end{itemize}
    \item The same operation can be repeated for multiple targets.
    \item  When doing add, you can also have volume and direction attributes
    \item When editing existing sound sources, you can also have mixed attributes in terms of the volume and sound directions.
    \item You must ensure each step logically contributes to the final transformation.
\end{itemize}

Return the output in the following structured JSON format:

\{
\begin{itemize}[label={}, itemsep=0.5pt, leftmargin=1.5em, topsep=0pt]
    \item "sound sources": ["...", "..."], here you should put the sound sources you are given
    \item  "complex editing instruction": "...",
    \item "atomic editing steps": [
    \begin{itemize}[label={}, itemsep=0.5pt, leftmargin=1.5em]
        \item \{"operation": "add", "target": "...", "effect": "at xxx(left, front, right) by xxxdB"\},
        \item \{"operation": "remove", "target": "...", "effect": "None"\},
        \item  \{"operation": "turn up/turn down", "target": "...", "effect": "xxxdB"\},
        \item \{"operation": "change", "target": "...", "effect": "to xxx (left, right, or front)"\},
    \end{itemize}
    ]
\end{itemize}
\}

Do NOT break down sound sources into individual words or characters. 
Sound sources are complete strings and must remain so in the JSON.
\end{gptprompt}

\newpage
\section{Implementation Details}
% only pure text is needed here
\subsection{Baselines implementation}
\label{app:baseline implementation}

We present the following baseline methods to evaluate the complex audio editing task. 
One of the baselines is an end-to-end version of Audit without using ALM. 
All other baseline methods perform multi-step sequential editing with ALM's atomic editing step outputs.

\header{End-to-End Audit.} 
We first train an end-to-end version of Audit that directly predicts the final-step edited audio $a_n$ conditioned on the original audio $a_0$ and high-level instruction $\mathcal{P}$ in one step without ALM. 
We extend the mono-channel Audit to our binaural setting, where we stack the left and right channels of the mel-spectrograms as the model inputs. 
Follow the original implementation, we convert 10s of audio into mel-spectrograms with a size of $80\times 624$. using a hop size of 256, a window size of 1024, and mel-bins of size 80.
We use the same model configurations in the original paper. 

\header{SDEdit.} 
SDEdit is a zero-shot method that does not require training a new audio editing model. It uses an off-the-shelf text-to-audio (TTA) generation model, which we use Stable-Audio-Open, as it supports binaural audio generation. 
We use the default 200 total diffusion steps and start the reverse process from a timestep of 100. We use a classifier-free-guidance (CFG) scale of 7.5. The target caption is composed by concatenating the individual event captions after each editing step. 

\header{DDIM Inversion.}
Similar to SDEdit, we also use Stable-Audio-Open as the TTA generation model. We use the default 200 total diffusion steps and start the reverse process from a timestep of 100. We use a CFG scale of 7.5 for both the target and the source. The source text prompt and the target text prompt are composed by concatenating the individual event captions before and after each editing step, respectively. 

\header{ZETA.}
Similar to SDEdit and DDIM Inversion, we also use Stable-Audio-Open as the TTA generation model. We use the default 200 total diffusion steps and start the reverse process from a timestep of 100. We use a CFG scale of 7.5 for the target and a CFG scale of 1 for the source. The source text prompt and the target text prompt are composed by concatenating the individual event captions before and after each editing step, respectively.

\header{AudioEditor.} 
Specifically in AudioEditor, we replace Affusion with the spatial generator SpatialSonic in BEWO to support binaural editing.
We use the default 100 total editing steps with 5 iterations per step to update text embedding for null-text inversion. 
For the addition task, we use the default punishment ratio (alpha) of -0.001. 
For the remove or extract task, we use the default punishment ratio (alpha) of 1.

\header{Audit with ALM.} 
To evaluate sequential editing with a trainable editor, we use an Audit baseline trained on single-step audio editing data with input audio $a_{i-1}$, atomic instruction $s_i$, and output edited audio $a_{i}$. 
The model and data configurations are the same with the end-to-end Audit baseline variant.

\subsection{\name{} implementation}
\label{app:smartdj implementation}

\header{ALM.} Our ALM module is initialized from Audio Flamingo 2.
This model contains an AF-CLAP audio encoder module to encode the mono-channel audio. 
The input 10-second audio is first resampled to 16KHz and transformed into a dense audio features $z_a\in \mathbb{R}^{64\times2048}$.
The mono-channel CLAP encoder understands the audio semantics and sound events, which is enough to reason the atomic steps to make the edited audio semantically align with the high-level instruction.
This audio encoder is followed by a representation transformation layers that expand the model capacity. 
This module has three self-attention layers to the audio feature representation, each with 8 heads and a dimension of 2048.
Following this, gated cross-attention layers are used to condition audio representations on the LLM. 
The LLM uses Qwen2.5-3B, a decoder-only causal LLM with 3B parameters, 36 hidden layers, and 16 attention heads.
During training, we keep both the AF-CLAP and LLM frozen during training.
The audio representation transformation layers are fully optimized.
We apply LoRA only to the gated cross-attention layers with a rank of 16.
% The ALM finetuning has a warm-up step of 2K.

\header{DiT architecture} We conduct experiments at 24KHz audio sample rate. The waveform latent representation operates at 50Hz and consists of 128 channels. 
Our DiT model has 24 blocks, each with a channel of 1024, 16 attention heads, and an FFN (Feed-Forward Network) dimension of 4096, resulting in total of 597M parameters.
During training, we begin with a 50K linear warm-up phase and the result is training with a cosine learning rate decay.
In the inference, we use 100 diffusion steps and a guidance scale of 4, with guidance rescale of 0.8.

\section{More results}
In this section, we provide more qualitative results on the ALM-inferenced editing steps and the LDM edited audio results.
\subsection{Results on ALM-inferenced atomic editing steps}
\label{app:alm_inference_results}
We show multiple examples on ALM-inferenced atomic editing steps.
In each example, the "audio labels" are the ground truth audio labels for reference and they are not provided to the ALM module.
Only the corresponding audios are provided to the ALM.
We then input the high level instruction to the ALM and it output the atomic editing steps.

\begin{example}{Example 1: Make this sound like a quiet afternoon in a garden}
Audio labels: clock tick, bird chirp, wind

High-level instructions: Make this sound like a quiet afternoon in a garden

Atomic editing steps:
 \begin{itemize}[label=-, itemsep=0.5pt, leftmargin=2em, topsep=0pt]
      \item Remove the sound of clock tick
      \item Turn up the the sound of bird tweet by 3dB
      \item Add the sound of gentle breeze at front by 2dB
\end{itemize}
\end{example}

\begin{example}{Example 2: Make this sound like a busy city street}
Audio labels: engine rev, church bell ring

High-level instructions: Make this sound like a busy city street

Atomic editing steps:
 \begin{itemize}[label=-, itemsep=0.5pt, leftmargin=2em, topsep=0pt]
      \item Remove the sound of bell ring
      \item Turn down the sound of engine rev by 2dB
      \item Add the sound of distant siren at left by 2dB
      \item Add the sound of traffic noise at front by 3dB
\end{itemize}
\end{example}

\begin{example}{Example 3: Make this sound like a bustling park on a sunny day.}
Audio labels: children scream, insect buzz, bird call, chainsaw run

High-level instructions: Make this sound like a bustling park on a sunny day.

Atomic editing steps:
 \begin{itemize}[label=-, itemsep=0.5pt, leftmargin=2em, topsep=0pt]
    \item Remove the sound of chainsaw run
    \item Add the sound of laughter at left by 3dB
    \item Turn down children scream by 2dB
    \item Change the sound of bird call to front
\end{itemize}
\end{example}

\begin{example}{Example 4: Make this sound like a serene beach}
Audio labels: water waves, whistling, music playing

High-level instructions: Make this sound like a serene beach

Atomic editing steps:
 \begin{itemize}[label=-, itemsep=0.5pt, leftmargin=2em, topsep=0pt]
      \item Remove the sound of whistling
      \item Turn up the sound of wave crash by 4dB
      \item Add the sound of seagulls calling at front by 3dB
\end{itemize}
\end{example}

\begin{example}{Example 5: Make this sound like a busy office}
Audio labels: typewriter tap, power tool drill

High-level instructions: Make this sound like a busy office

Atomic editing steps:
 \begin{itemize}[label=-, itemsep=0.5pt, leftmargin=2em, topsep=0pt]
      \item Remove the sound of drilling
      \item Turn up the sound of typewriter type by 2dB
      \item Add the sound of phone ringing at right by 3dB
\end{itemize}
\end{example}

\begin{example}{Example 6: Make this sound like a military base}
Audio labels: fire truck siren, gunshot, vehicle move

High-level instructions: Make this sound like a military base

Atomic editing steps:
 \begin{itemize}[label=-, itemsep=0.5pt, leftmargin=2em, topsep=0pt]
      \item Remove the sound of fire truck siren
      \item Turn up the sound of gunshot by 3dB
      \item Add the sound of helicopter blades at right by 2dB
\end{itemize}
\end{example}

\begin{example}{Example 7: Make this sound like a calm night in the countryside}
Audio labels: snoring, machine run, wind blow

High-level instructions: Make this sound like a calm night in the countryside

Atomic editing steps:
 \begin{itemize}[label=-, itemsep=0.5pt, leftmargin=2em, topsep=0pt]
      \item Remove the sound of machine run
      \item Turn down the sound of wind blow by 2dB
      \item Add the sound of owl hooting at right by 2dB
\end{itemize}
\end{example}

\begin{example}{Example 8: Make this sound like a bustling urban street}
Audio labels: horn honk, cat, siren blare, male laugh

High-level instructions: Make this sound like a bustling urban street

Atomic editing steps:
 \begin{itemize}[label=-, itemsep=0.5pt, leftmargin=2em, topsep=0pt]
    \item Remove the sound of cat
    \item Turn down the sound of siren blare by 2dB
    \item Add the sound of people chatter at front by 3dB
    \item Add the sound of traffic noise at left by 2dB
\end{itemize}
\end{example}

\subsection{Results on atomic editing steps}
\label{app:edit_examples}

\begin{figure}[!h]
    \centering
    \includegraphics[width=\linewidth]{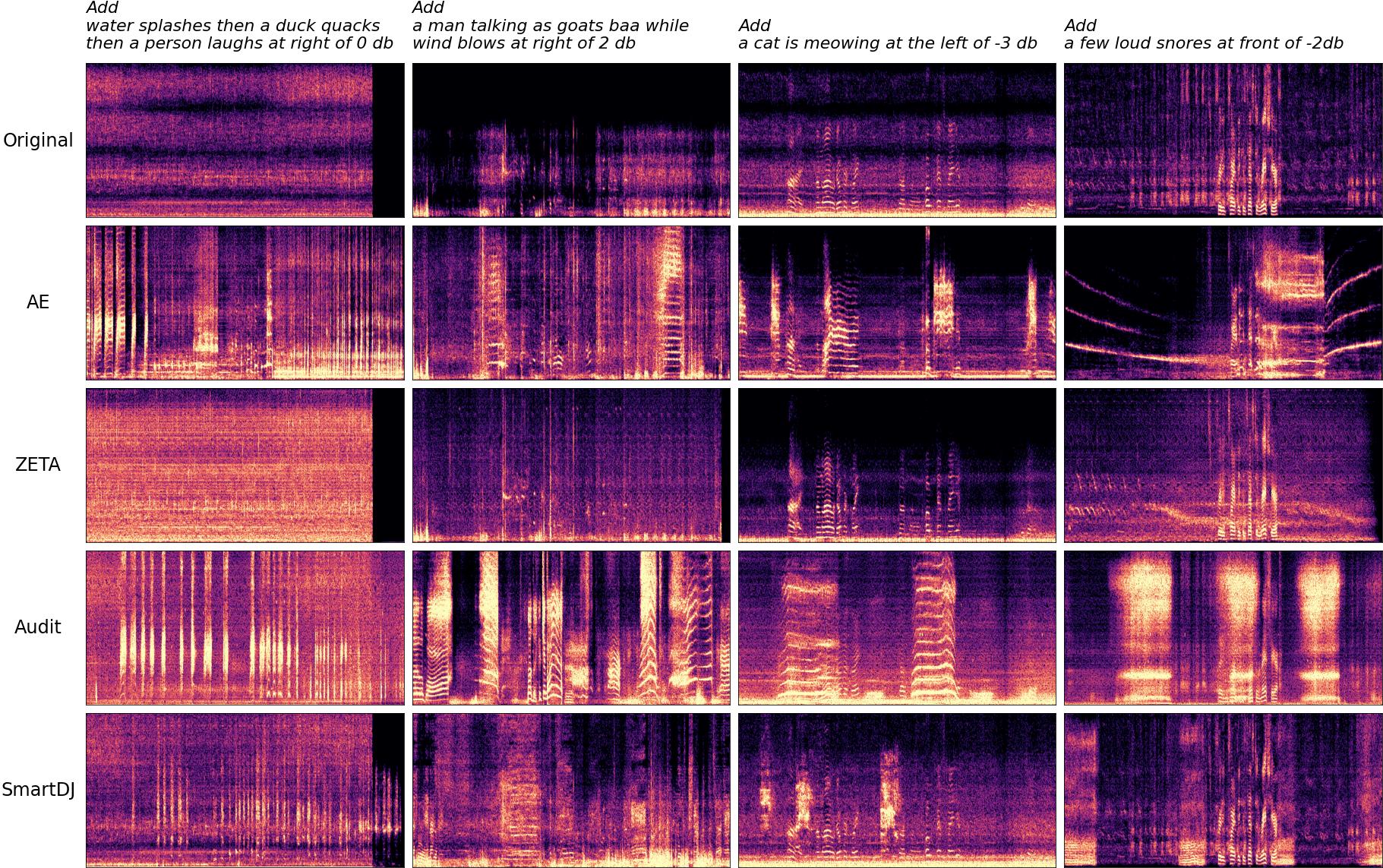}
    \caption{Examples on \textit{add} operation. 
    The top row is the original audio and the rest rows are the edited results. Only \name{} can keep the original audio clips while add new audio contents.}
    \label{fig:add_example_more}
\end{figure}

We provide more atomic editing results on \textit{add} in Fig~\ref{fig:add_example_more}.
As shown by the comparison with the original audio and the edited one, while baseline method tends to replace the original clips and completely generate a new audio clip, \name{} can successfully keep the original contents and \textit{add} new sound events into it.

\begin{figure}[!h]
    \centering
    \includegraphics[width=\linewidth]{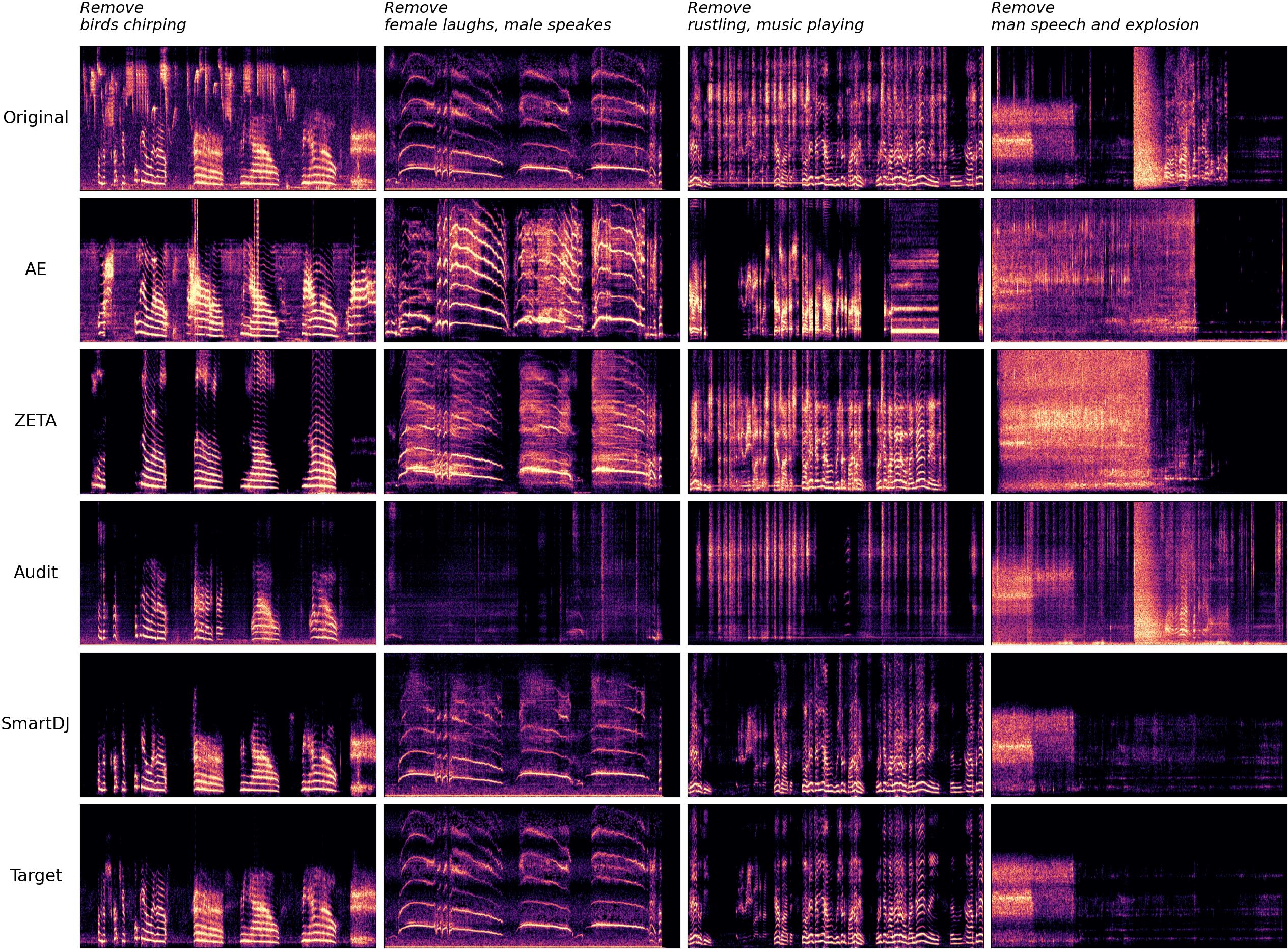}
    \caption{Examples on \textit{remove} operation. The top row is the original audio and the bottom row is the target audio. Only \name{} can completely remove unwanted audios parts and keep the remaining part unchanged.}
    \label{fig:remove_example_more}
\end{figure}

\begin{figure}[!h]
    \centering
    \includegraphics[width=\linewidth]{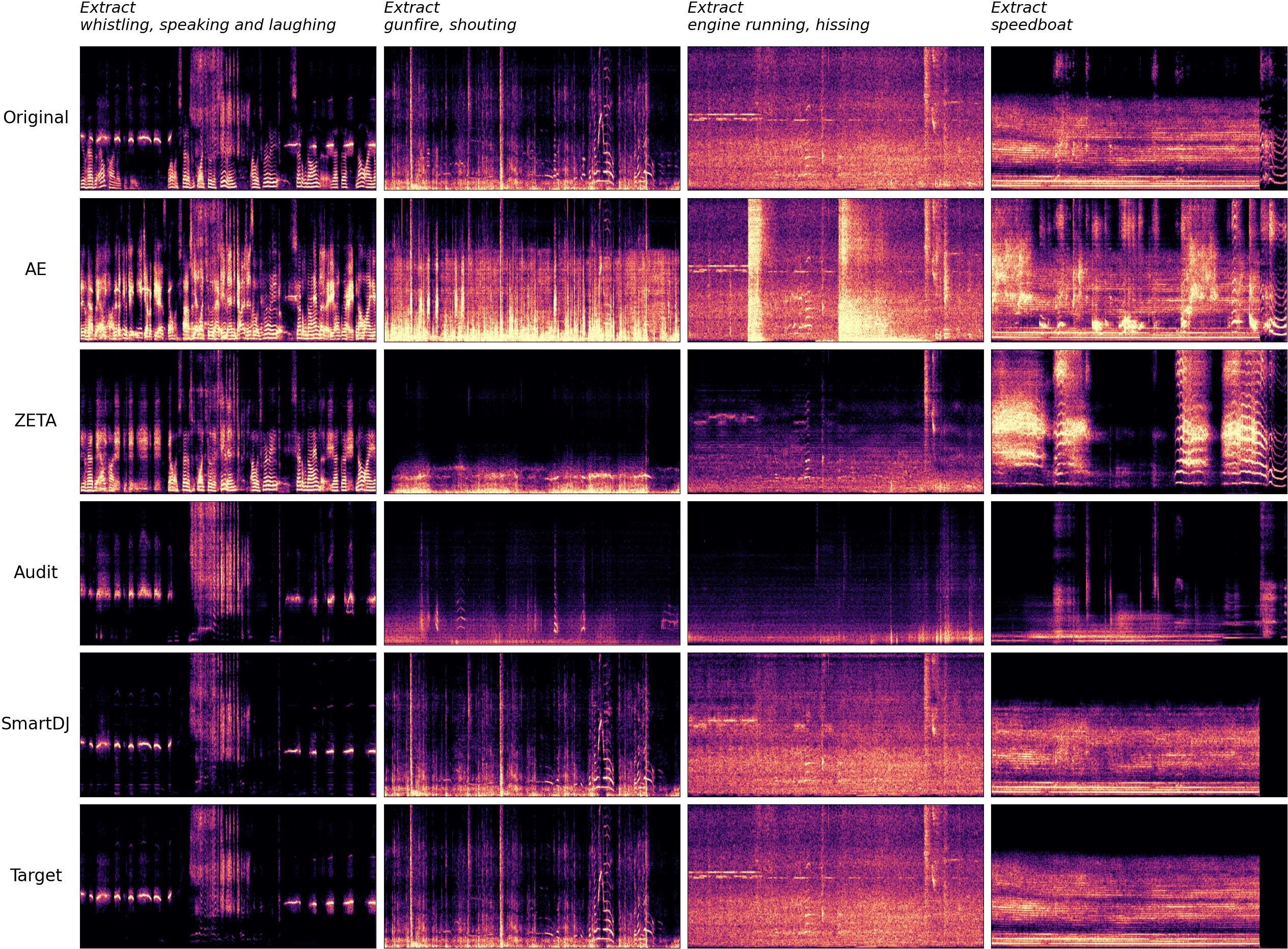}
    \caption{Examples on \textit{extract} operation. The top row is the original audio and the bottom row is the target audio. Only \name{} can extract wanted audios that are clean and of high quality.}
    \label{fig:extract_example_more}
\end{figure}

We show more editing examples on \textit{remove} and \textit{extract} in Fig~\ref{fig:remove_example_more} and ~\ref{fig:extract_example_more}.
Compared with baseline method, \name{} can effectively either remove unwanted or extract wanted audio contents.
The edited results show good alignment with the ground truth edited audios.

\begin{figure}[!h]
    \centering
    \includegraphics[width=\linewidth]{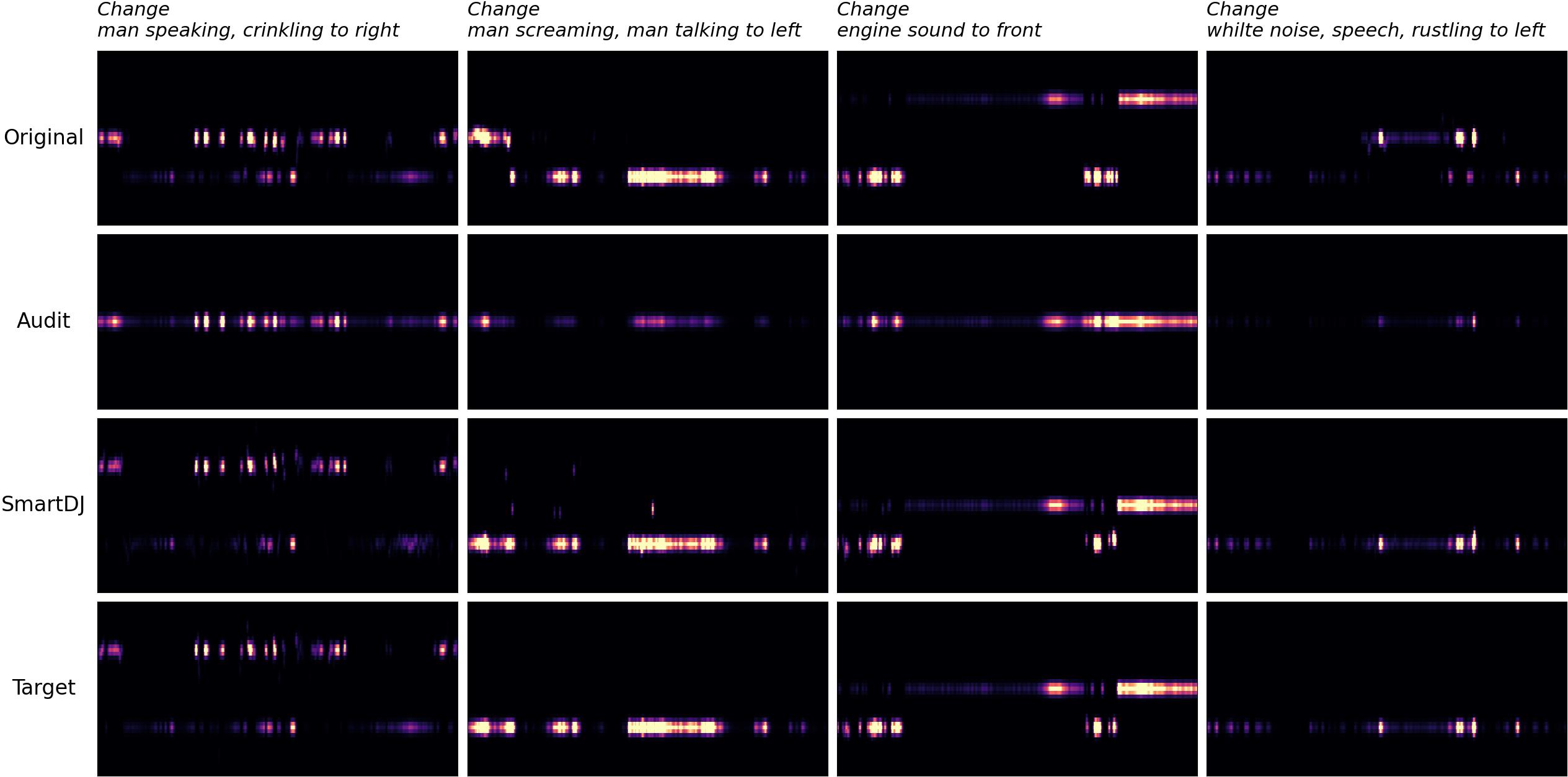}
    \caption{Examples on \textit{change} operation. The top row is the original audio and the bottom row is the target audio. Only \name{} can perfectly edit the sound directions that are matching closely with the target.}
    \label{fig:change_example_more}
\end{figure}

Figure \ref{fig:change_example_more} presents additional qualitative results for the \textit{change sound direction} task. 
The y axis in each figure is the sound direction heatmap.
\name{} consistently relocates the source to the requested spatial direction, and its outputs align well with the ground truth edited audios.
By contrast, Audit can not alter spatial effect, showing the limitations of using spectrogram audio encoder for direction-aware editing.

\newpage

\begin{figure}[!htp]
  \centering

  \begin{subfigure}{0.52\textwidth}
    \centering
    \includegraphics[width=\linewidth]{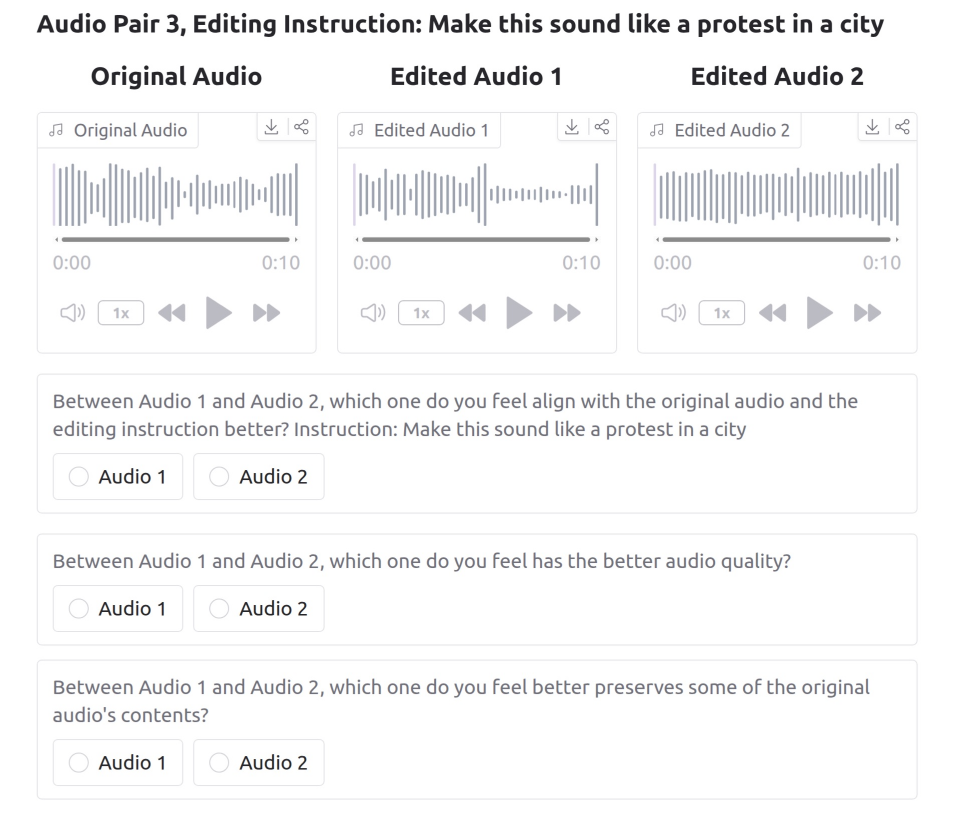}
    \caption{User interface for complex audio editing task}
    \label{fig:user_study_interface_1}
  \end{subfigure}
  \hfill
  \begin{subfigure}{0.47\textwidth}
    \centering
    \includegraphics[width=\linewidth]{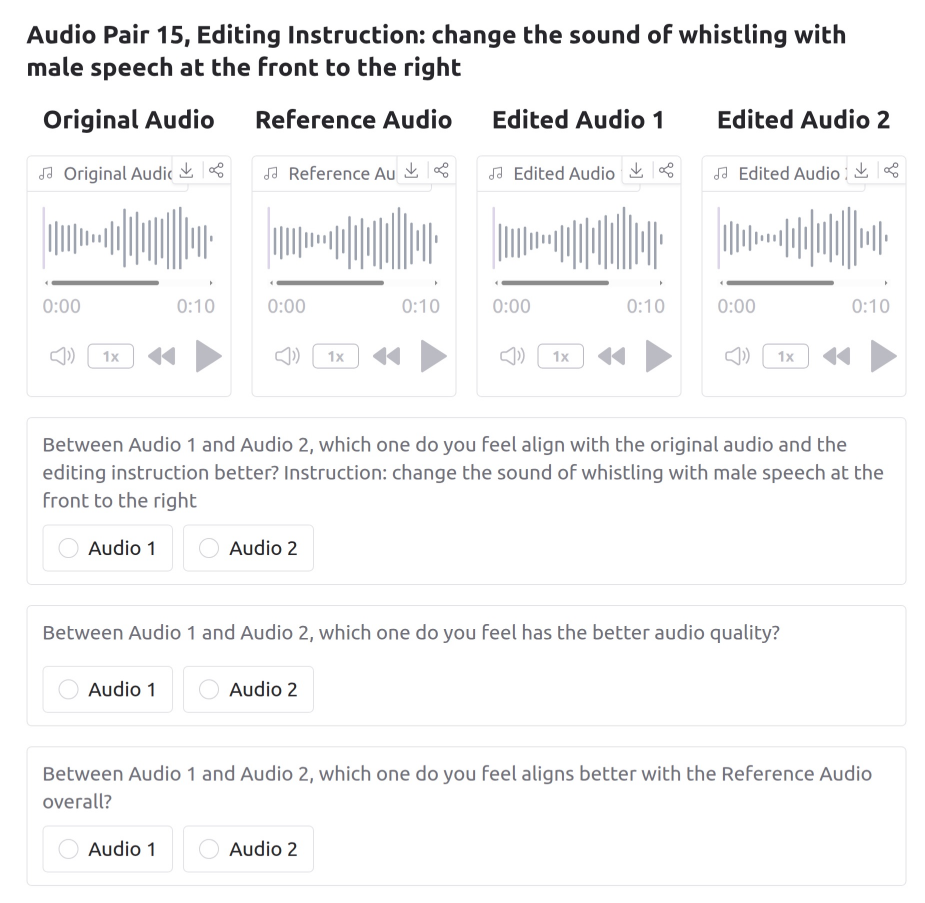}
    \caption{User interface for change direction editing task}
    \label{fig:user_study_interface_2}
  \end{subfigure}

  \caption{The audio pairs, questions, and user interfaces for different audio editing tasks}
  \label{fig:user_study_interface}
  \vspace{-10pt}
\end{figure}

\subsection{Human Subjective Studies Details}
\label{app:human_subject_results}
We provide more details on the subjective user study.
We provide users with data pairs consisting of the original audio, the editing instruction, the \name{} edited result, and the edited results from a random competing baseline. 
In the case of single-step audio editing tasks \textit{remove, extract, turn up/down, change sound direction} where there exists a ground-truth editing solution, we also provide it as the reference audio as shown in Fig. \ref{fig:user_study_interface_2}. 
We conduct evaluations with 19 participants and 20 (10 complex audio editing, 10 single-step editing) data pairs per participant.

In the 10 complex audio editing question pairs, we ask the user to select between the \name{} edited audio and the edited results from a random competing baseline (randomly sampled from AudioEditor, ZETA and Audit) according to the following three questions:

\begin{question}{Question list for complex audio editing user study}
 \begin{itemize}[label=-, itemsep=0.5pt, leftmargin=2em, topsep=0pt]
    \item Between Audio 1 and Audio 2, which one do you feel aligns with the original audio and the editing instruction better?
    \item Between Audio 1 and Audio 2, which one do you feel has the better audio quality?
    \item Between Audio 1 and Audio 2, which one do you feel better preserves some of the original audio's contents?
\end{itemize}
\end{question}

In the single-step editing for \textit{add} operation (3 pairs in total), we compare \name{} with a randomly sampled method from three baselines (AudioEditor, ZETA and Audit). 
We ask the user to answer four questions, with one additional question listed below:

\begin{question}{Additional question for \textit{add} operation user study}
 \begin{itemize}[label=-, itemsep=0.5pt, leftmargin=2em, topsep=0pt]
    \item Between Audio 1 and Audio 2, which one do you feel the added spatial effect aligns with the text instruction?
\end{itemize}
\end{question}

For the single-step \textit{remove} and \textit{extract} tasks (4 pairs) we compare \name{} with a randomly chosen baseline model.
For \textit{turn-up/turn-down} and \textit{change direction} (3 pairs) we benchmark only against \textsc{Audit}, since the other baselines cannot perform these operations.
Each pair includes a ground-truth reference, and listeners answer the three evaluation questions listed below.

\begin{question}{Questions list for the \textit{Remove, Extract, Turn up/down, Change sound direction} operation user study}
 \begin{itemize}[label=-, itemsep=0.5pt, leftmargin=2em, topsep=0pt]
    \item Between Audio 1 and Audio 2, which one do you feel aligns with the original audio and the editing instruction better?
    \item Between Audio 1 and Audio 2, which one do you feel has the better audio quality?
    \item Between Audio 1 and Audio 2, which one do you feel aligns better with the Reference Audio overall?
\end{itemize}
\end{question}

Across all tasks, \name{} is consistently preferred over all baseline methods.
For complex audio editing task, \name{} receives at least 80\% of user votes over the baselines for audio quality, and at least 87\% for alignment with the high-level editing instruction and original audio.
This shows that \name{} faithfully performs the requested scene transformation while preserving the key elements of the original audio.
In the single-step editing task, our method receives more than 77\% of user votes for audio quality, and 84\% for alignment with the single-step editing text prompt, spatial description, and original or reference audio (when applicable).
These results demonstrate that \name{} achieves the highest user preference across both quality and alignment metrics, outperforming all competing methods.